\documentclass[superscriptaddress,floatfix,showpacs,aps,prb,numerical,twocolumn]{revtex4-1}
\usepackage{graphicx}
\usepackage{color}
\usepackage{natbib}
\usepackage{amssymb,amsmath}
\usepackage{textcomp}
\usepackage{multirow}\newcommand{\dg}{\,^{\circ}\mathrm{C} }
 
\renewcommand{\vec}[1]{\mathbf{#1}}

\begin{document}
\title{On the analysis of island shape evolution from diffuse x-ray scattering of organic thin films and the implications for growth}
%
\author{C. Frank}
 \affiliation{Institut f\"{u}r Angewandte Physik, Universit\"{a}t T\"{u}bingen, Auf der Morgenstelle 10, 72076 T\"{u}bingen, Germany}

\author{R. Banerjee}
 \affiliation{Institut f\"{u}r Angewandte Physik, Universit\"{a}t T\"{u}bingen, Auf der Morgenstelle 10, 72076 T\"{u}bingen, Germany}

\author{M. Oettel}
 \affiliation{Institut f\"{u}r Angewandte Physik,  Universit\"{a}t T\"{u}bingen, Auf der Morgenstelle 10, 72076 T\"{u}bingen, Germany}  

\author{A. Gerlach}
 \affiliation{Institut f\"{u}r Angewandte Physik, Universit\"{a}t T\"{u}bingen, Auf der Morgenstelle 10, 72076 T\"{u}bingen, Germany} 

\author{J. Nov\'{a}k}
 \affiliation{Central European Institute of Technology, Masaryk University, Kamenice 5, CZ-62500 Brno, Czech Republic}  
 
\author{G. Santoro}
 \affiliation{Photon Science, DESY, Notkestr. 85, 22607 Hamburg, Germany} 
   
\author{F. Schreiber}
 \affiliation{Institut f\"{u}r Angewandte Physik, Universit\"{a}t T\"{u}bingen, Auf der Morgenstelle 10, 72076 T\"{u}bingen, Germany} 

\date{\today}
\begin{abstract}
Understanding the growth of organic semi-conducting molecules with shape anisotropy is of high relevance to the processing of optoelectronic devices. This work provides insight into the growth of thin films of the prototypical rodlike organic semiconductor diindenoperylene on a microscopic level, by analyzing in detail the film morphology. We model our data, which were obtained by high-resolution grazing incidence small angle x-ray scattering (GISAXS), using a theoretical description from small angle scattering theory derived for simple liquids. Based on form factor calculations for different object types we determine how the island shapes change in the respective layers. Atomic force microscopy measurements approve our findings.  
\end{abstract}
%
\maketitle
\section{Introduction}

The investigation of growth processes and related changes in the interface morphologies are extremely relevant in many scientific areas. One of the ideally suited experimental methods to study the kinetic effects involved, such as surface diffusion, island condensation, and island nucleation~\cite{TMichely_2004_book,Fendrich_2007_PhysRevB}, all of which are inherently connected with the growth process itself, is x-ray scattering~\cite{Sinha_1994_PhysicaB,Woll_2011_PhysRevB,Kowarik_2008_PhysStatusSolidiRRL}. Particularly, diffuse scattering techniques~\cite{Sinha_1996_PhysicaA,Pershan_2000_COLLOIDSURFACEA,Nickel_2004_PhysRevB} have widely been employed to decipher such processes. Apart from the surface correlations~\cite{Sinha_1988_PhysRevB,Tolan,Salditt_1994_PhysRevLett}, \textit{in situ} studies allow to monitor the growth and the evolution of the surface morphology in real time~\cite{Fleet_2006_PhysRevLett, Frank_2014_dip-realtime,Yu_JPhysChemLett_2013,Gonzalo_APL_2014}.   

In contrast to grazing incidence diffraction (GID), where the in-plane lattice planes of the crystallites are probed~\cite{Frank_2013_JApplPhys} on a molecular level, grazing incidence small angle x-ray scattering (GISAXS) provides access to length-scales ranging from several tens of nanometers to $\simeq 1 \mu$m~\cite{Lazzari_2007_PhysRevB,Buschbaum_2009_Book}. Therefore, among the prevalent off-specular scattering techniques, GISAXS is the ideal tool to characterize the morphology of the sample, while simultaneously yielding a complete and non-invasive, statistical averaging of the surface (within the limits of the transverse coherence length of the X-ray beam)~\cite{Smilgies_2002_SynchrotronRadiationNews,Metzger_1999_JPhysDApplPhys,Du_2004_AdvMater}. However, a quantitative analysis of the GISAXS data from island sizes, island-island correlations, and island shapes can require a significant computational and numerical effort. Although such kind of analysis has successfully been employed for well-ordered inorganic 2D structures~\cite{Stangl_1999_APL}, a generalization to organic materials, particularly to those with steps in the morphology and with shape anisotropy on the molecular level, is to our knowledge, still lacking.

In this study we intend to outline a general approach to the quantitative analysis of GISAXS data using the inverse Fourier transform of different island form factors.
As a representative material for rodlike organic semiconductors we use diindenoperylene (DIP, C$_{32}$H$_{16}$)~\cite{Kurrle_2008_APL,deOteyza_2009_afm,deOteyza_2007_apl,WagnerAM10}, which is a crystalline small-molecule with significant potential for optoelectronic devices~\cite{Brutting_2012_book,Tripathi_2006_ApplPhysLett}, due to the large hole mobility~\cite{Tripathi_2006_ApplPhysLett}, ambipolar charge carrier transport in donor:acceptor blends~\cite{OpitzIEEE}, as well as interesting structural properties in the pure~\cite{Hinderhofer_2012_ApplPhysLett,Kowarik_2009_AEM,Zhang_2009_PhysRevLett,Durr_2003_PhysRevLett,Durr_2005_ThinSolidFilms,HeinrichJPCC07,Kowarik_2006_PhysRevLett} and mixed phases~\cite{Banerjee_2013_PhysRevLett,Aufderheide_2012_PhysRevLett}.
 
We provide a combined GISAXS and atomic force microscopy (AFM) study on a set of \textit{ex situ} samples with film thicknesses covering the first few monolayers, which represent the initial stages of the growth. Using a theoretical description from small angle scattering, we model our data with the form factor in the Born approximation (BA), taking different island shapes into account. The Born approximation form factor can provide a good approximation when both, the incident angle and the exit angle are higher than the critical angle. In this angular regime the dominant term of the distorted wave Born approximation (DWBA) corresponds well to the BA and multiple scattering and absorption effects can be neglected.~\cite{Revenant_PRB_2004,Daillant_2009_book} For computational efficiency in the island shape fits, we have solved the Hankel transform for the in-plane component of the momentum transfer using the logFFT algorithm. Subsequently, we compare the estimated island size with that, obtained from AFM measurements. Our results are qualitatively corroborated by Ref.~[\onlinecite{Frank_2014_dip-realtime}], in which the growth kinetics and the nucleation behavior of ultra-thin films of DIP were investigated in real-time, thus allowing for the determination of the mean island size and the molecular diffusion in the very first layers.

Based on our results, we finally suggest a model on how the average shape of the islands changes in the respective layers as the growth progresses.
We emphasize that our experimental strategy as well as our approach for the data analysis should be equally applicable to other systems with evolution of islands during growth.

\section{Experimental}
\begin{figure}[tpb]
\centering
\includegraphics[width=0.45\textwidth]{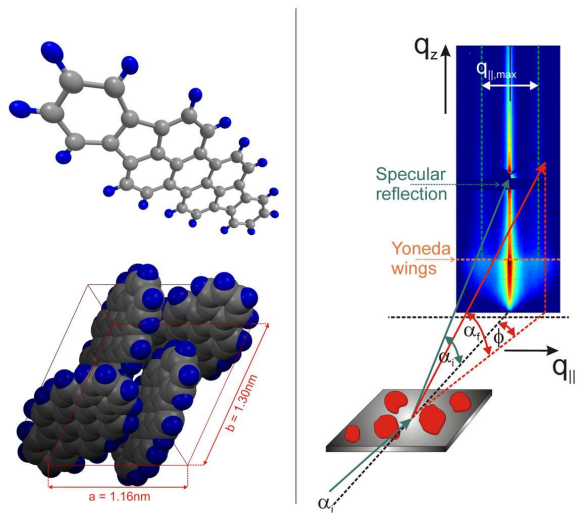}
\caption{(Color online) Left panel: (top) Single DIP (C$_{32}$H$_{16}$) molecule shown from a perspective view and (bottom) DIP molecules (at 298\,K) arranged in the unit cell with the view along the c*-axis. The atom positions were taken from Ref.~[\onlinecite{HeinrichJPCC07}]. Right panel: GISAXS scattering geometry. The specular reflection is shielded with a beamstop, while the diffuse intensity, which is caused by the surface roughness, is recorded in the reciprocal space map (RSM) as a function of the exit angle $\alpha_f$ and the in-plane angle $\phi$.}
\label{fig:gisaxs_geometry}
\end{figure}
Using an ultrasonic bath, the SiO$_x$ substrates were cleaned with acetone, iso-propanol and ultra-pure water. DIP (see Fig.~\ref{fig:gisaxs_geometry}) was purchased (with gradient sublimation purity) from the Institute f\"ur PAH Forschung (Greifenberg, Germany). Samples were prepared in ultra-high vacuum (UHV) conditions using the organic molecular beam deposition (OMBD) technique akin to the procedure described, e.g.\ in Ref.~[\onlinecite{Ritley_2001_RevSciInstrum, Witte_2004_JMaterRes, Schreiber_2004_Physstatsola, Kowarik_2008_PhysStatusSolidiRRL,Hinderhofer_2012_ChemPhysChem}].

To be able to perform a consistent and thorough data analysis, several DIP samples were grown with film thicknesses (or equivalently total film coverages) ranging from $\sim$\,0.5\mbox{--}5.5\,ML. Notably, to obtain a strong diffuse scattering intensity we have chosen the respective layer coverages of the films in fractions corresponding to half a monolayer~\cite{Frank_2014_dip-realtime}.
The growth rate for this sample series was set to $R_\mathrm{growth}\approx 0.1$\,nm/min and the substrate temperature was fixed to $\simeq 25\,\dg$.

In order to investigate how the in-plane morphology of DIP changes as a function of deposited material, GISAXS measurements (see Fig.~\ref{fig:gisaxs_geometry} for schematic) were performed at the P03 MiNaXS beamline~\cite{Roth_JPhysCondensMatter_2011,Buffet_JSynRad_2012,Santoro_RevSciInstrum_2014} at the PETRA III storage ring, DESY, which is ideally suited to measure long-range in-plane correlation lengths. In the experiment, the detector-to-sample distance was 4.9\,m and a Pilatus 300K detector with a pixel size of $172\times 172$\,\textmu m$^2$ was used. 
During the measurements the incidence angle was set to $\alpha_i = 0.39^{\circ}$ and a wavelength of $\lambda = 1.0868$\,\AA{} was employed. 
In order to minimize the effect of beam damage on our samples, synchrotron measurements were performed in a controlled environment provided by a chamber purged with nitrogen and equipped with kapton windows. For each of the samples 51 frames were recorded with an exposure time of one second per frame. No change could be observed within the series of frames corresponding to the same sample, hence we exclude a significant impact of beam damage. The respective frame series were, finally, binned to one image leading to an improved signal in the GISAXS images.

Our AFM measurements were performed in non-contact mode on a JPK Nanowizard II instrument. The AFM data analysis was performed using the software Gwyddion~\cite{Necas_2011_centreurjphys}.

\section{Results and Discussion}

\subsection{Discussion of the reciprocal space maps}
\begin{figure*}[tpb]
\centering
\includegraphics[width=0.8\textwidth]{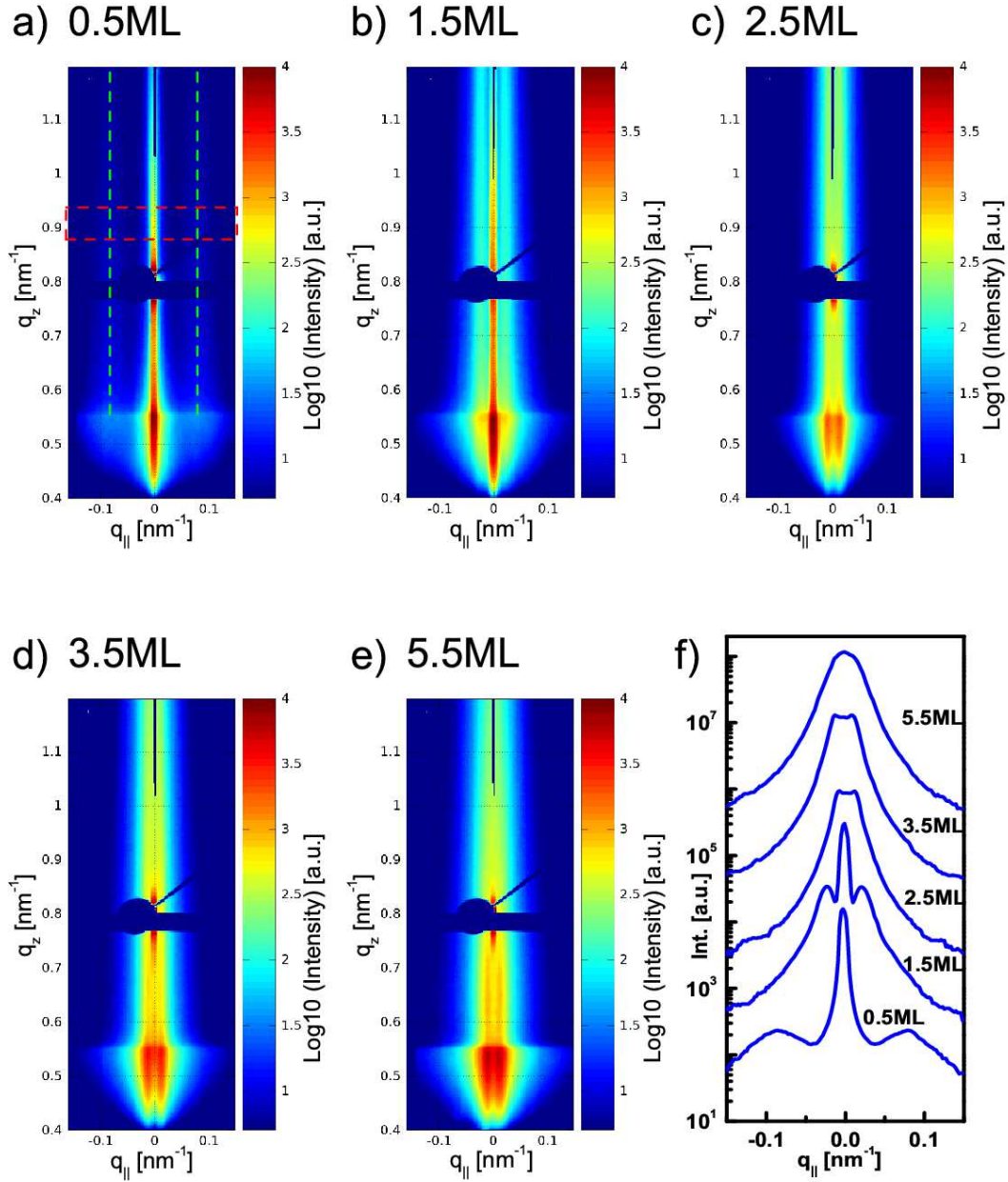}
\caption{(Color online) (a-e) GISAXS signal of ultra thin films of pure DIP grown on native SiO$_x$ substrates. The film thickness covers a range of 0.5\mbox{--}5.5\,ML. The dark blue horizontal stripe corresponds to the non-sensitive inter module detector gaps, while the dark circle corresponds to the specular beamstop. Green dashed lines in (a) indicate the positions of the correlation streaks. The red box in (a) shows at which $q$-position the region of interest (ROI) was chosen in order to extract the GISAXS line profiles. To improve the statistics of the line profiles the ROI was integrated along the $q_z$-direction. The resulting horizontal GISAXS sections are shown for the respective coverages in (f). One observes that the maximum position of the correlation peaks shifts towards smaller $q_{||}$ for increased film thickness indicating the increase in the in-plane correlation length as a function of film thickness. Note that the curves have been shifted for clarity in (f).}
\label{fig:desy_gisaxs}
\end{figure*}
Although, in principle AFM gives similar information as GISAXS, there are situations were the AFM technique is not ideal. Apart from the fact that the AFM is primarily a local probe and the images obtained are not necessarily a true representative of the average topography, there is always the risk of tampering the top organic layers particularly, when measurements have to be performed \textit{in situ} at a sufficiently short time scale~\cite{Frank_2014_dip-realtime}. Therefore, in order to test the applicability of our method in real-time the data analysis is first tested under conditions, which can be compared to AFM. Accordingly, we employ GISAXS as the main technique in this study.

Figure~\ref{fig:desy_gisaxs} shows the GISAXS signal as reciprocal space maps (RSM) for the binned images covering a film thickness between 0.5\mbox{--}5.5\,ML.  
The position $q_{z,c}$ of the Yoneda wing~\cite{Sinha_1988_PhysRevB} was extracted and translated into the critical angle $\alpha_c$ according to
\begin{center}
  \begin{align}
      \alpha_c=\sin^{-1}\left(\frac{q_{z,c}\lambda}{2\pi}-\sin\alpha_i\right)\, , 
  \end{align}
\end{center}
 leading to $\alpha_c(\mathrm{SiO_x})\approx 0.163^{\circ}$ and $\alpha_c(\mathrm{DIP})\approx 0.15^{\circ}$ (see also Ref.~[\onlinecite{Durr_2003_PRB, Durr_2002_APL}]. Note that including dispersion $\delta$ and absorption  $\beta$ of the x-rays, the penetration depth~\cite{Daillant_2009_book} $z$ is given by
 \begin{center}
  \begin{align}
 z_{1/e}=-\lambda/(4\pi B(\alpha_i))\, 
 \end{align}
\end{center}
with
 \begin{center}
  \begin{align}
      B(\alpha_i)=-\frac{1}{\sqrt{2}}\sqrt{\sqrt{(\alpha_i^2-2\delta)^2+4\beta^2}-(\alpha_i^2-2\delta)}\, .
  \end{align}
\end{center}
Therefore, under the specular condition of reflection, i.e.\ where $\alpha_i = \alpha_f = 0.39^{\circ}$ and correspondingly $q_z = 0.79$\,nm$^{-1}$, the penetration depth of the incoming wave is of the order of $\sim 1.25\times 10^4$\,\AA{}, resulting in a full penetration of the thin organic layer.   

A further prominent scattering feature in GISAXS experiments is the presence of side streaks along the $q_z$-direction in the RSM. These (in the following called) ``correlation streaks'' point to a certain in-plane correlation length, which is caused by lateral roughness modulations in the morphology of the sample. In all of the images of Fig.~\ref{fig:desy_gisaxs} we find two such streaks at different $q_{||}$-positions symmetrically located around $q_{||}=0$. It is observed that the $q_{||}$-positions of the streaks strongly depend on the film thickness. In particular, we find that for larger thicknesses the separation between the two streaks decreases. Since we are mainly interested in the in-plane component of the momentum transfer, line profiles were extracted from the images in Fig.~\ref{fig:desy_gisaxs}. 
By choosing a suitable region of interest (ROI) as indicated by the red box in Fig.~\ref{fig:desy_gisaxs}(a), in which the intensity was integrated along the $q_z$-direction (in the range $q_z = 0.88 - 0.94$\,nm$^{-1}$) we get the horizontal GISAXS sections only as a function of $q_{||}$ as shown in Fig.~\ref{fig:desy_gisaxs}(f)
for the different film thicknesses.

\begin{figure}
 \centering
 \includegraphics[width=0.36\textwidth]{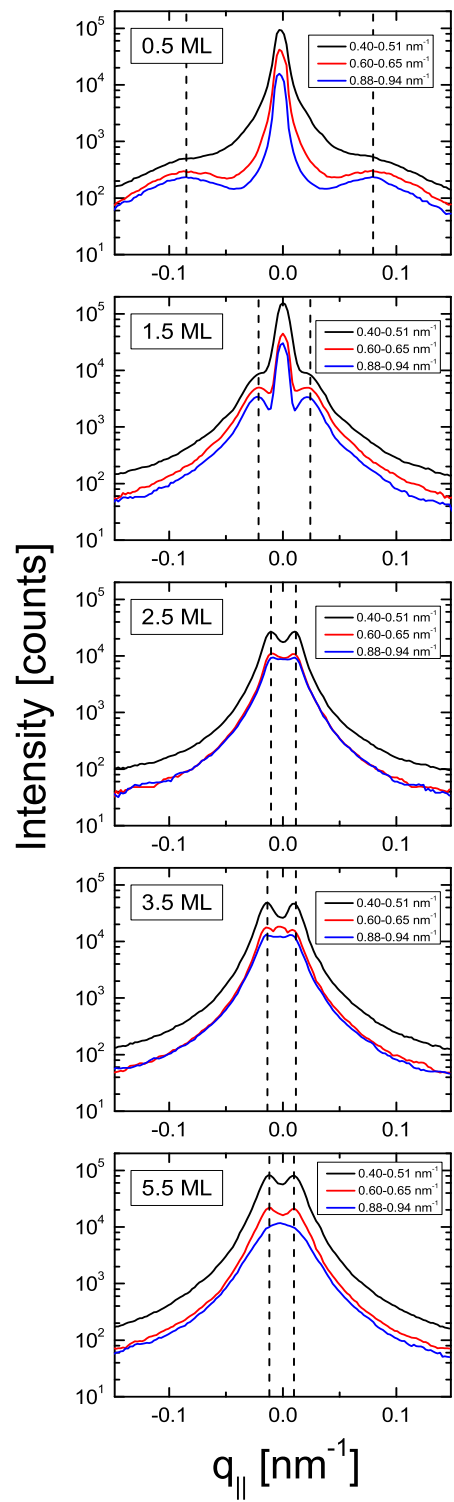}
 \caption{(Color online) Horizontal sections of the GISAXS-signal taken from Fig.~\ref{fig:desy_gisaxs}. Sections are extracted in the ranges $q_{z,1} = 0.4\mbox{--}0.51$\,nm$^{-1}$, $q_{z,2} = 0.6\mbox{--}0.65$\,nm$^{-1}$ and $q_{z,3} = 0.88\mbox{--}0.94$\,nm$^{-1}$ and shown for thicknesses between $0.5\mbox{--}5.5$\,ML. Dashed lines indicate that the position of the correlation peak does not change with respect to $q_z$.} 
 \label{fig:DESY_horsectb}
\end{figure}

For small thicknesses the samples exhibit a very pronounced central peak at $q_{||} = 0$. Due to the increasing film roughness, which usually occurs during the deposition of more DIP material~\cite{Durr_2003_PhysRevLett}, this peak gradually decreases and is finally masked by the approaching correlation peaks. 
Importantly, and relevant for the analysis in the following sections, we find that the correlation peak position in the line profiles (or correlation streaks in the RSM) do not show any dependence on $q_z$. This is observed for all thicknesses and is illustrated in Fig.~\ref{fig:DESY_horsectb} for the respective samples. As a consequence, the horizontal line profiles can be extracted at arbitrary $q_z$-positions within the GISAXS images, while still containing enough information (particularly, at $q_z$-positions well above the Yoneda wings, where multiple scattering and absorption effects are essentially negligible). As a consequence, the Born approximation~\cite{Daillant_2009_book} still holds and an appropriate theo\-retical model can be used in the data analysis. Albeit, there are pre-built tools for GISAXS data analysis, such as the software package ``FitGISAXS''~\cite{Babonneau_2010_JApplCrystallogr}, which is based on the distorted wave Born approximation, we find that some of the more realistic island shapes, e.g.\ a cone, are currently not supported. Therefore, we restrict ourself to model the line profiles with the description provided in the next section.

\subsection{Modeling the GISAXS profiles}
In the following, we introduce a simple model similar to Refs.~[\onlinecite{Lee_2003_ApplPhysLett,Basu_phys_rep_2002,Pietsch_2004_book}] to describe the scattering intensity $I$. In this approach, the scattering occurs from $N$ identical ``objects'', i.e.\ islands, with a 
three--dimensional shape, which are distributed on a two--dimensional plane (parallel to the substrate) such that their distribution
only depends on the in--plane coordinates. The average two--dimensional density of the islands is denoted by $\rho$. 
In general, we decompose the three--dimensional scattering vector $\vec q$
 and the three--dimensional position vector $\vec r$ into in--plane ($||$) and $z$--components:
\begin{align}
 \centering
 \vec{q}=\begin{pmatrix} \vec{q_{||}} \\ q_z \end{pmatrix},\hspace{0.5cm} \vec{r}=\begin{pmatrix} \vec{r_{||}} \\ z \end{pmatrix} 
 \label{eq:qvec}
\end{align}
 Hence the scattering intensity can in general be expressed as a product of an island form factor
and a factor describing the in--plane distribution of the islands~\cite{Stangl_1999_APL, Hansen_2006_book, guinier_1994} 
\begin{align}
\centering
\label{eq:Idiff_1}
I(\vec{q_{||}},q_z)&\propto |F(\vec q_{||}, q_z)|^2\left\langle\sum_{j=1}^N\sum_{k=1}^N e^{i\vec{q_{||}}\cdot(\vec r_{j,||}-\vec r_{k,||})}\right\rangle\, ,
\end{align}
where 
$\vec r_{j,||}$ is the in--plane position vector of island $j$ and the form factor $F$ is defined as the Fourier transform of the three--dimensional
island shape function $\Omega_s(\vec r_{||},z)$. 
Introducing the in--plane pair correlation function for the islands, $g(\vec r_{||})$, and the total correlation function 
$h(\vec r_{||})=g(\vec r_{||})-1$, one can define a two--dimensional structure factor by 
$S(\vec q_{||})= 1 +\rho \tilde h(\vec q_{||})$  
where $\tilde h(\vec q_{||}))={\rm FT}_{2D} h(\vec r_{||})$ is the two--dimensional Fourier transform of $h$.
With these definitions, the scattering intensity becomes
\begin{align}
\centering
I(\vec q_{||},q_z)\propto N |F(\vec q_{||}, q_z)|^2 \left( S(\vec q_{||}) + (2\pi)^2 \rho \delta^{(2)}(\vec q_{||})  \right)  \, .
\label{eq:Idiff2}
\end{align}
Thus $I$ contains a delta-peak corresponding to the in--plane forward direction. In the following, this peak will be neglected since it is not resolved
by the measured diffuse scattering intensity $I \equiv I_{\rm diff}$~\cite{Lazzari,guinier_1994,Renaud_2009_SurfSciRep}.

\begin{figure*}[htb]
 \centering
 \includegraphics[width=0.85\textwidth]{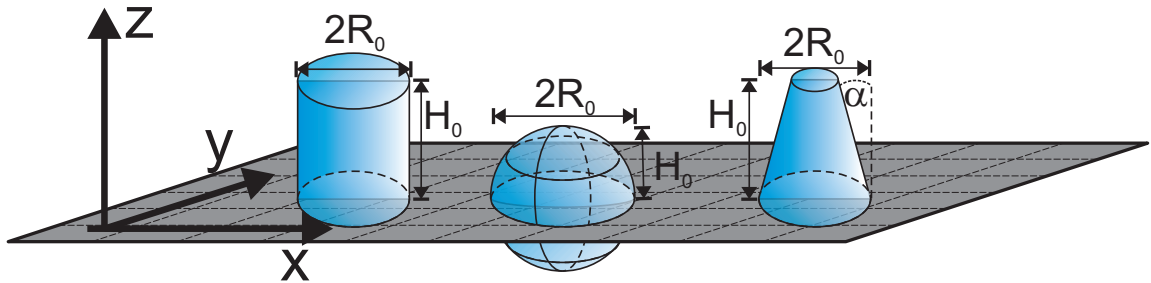}
 \caption{(Color online) Geometries of the island shape used to fit $\Phi$ in Eq.~(\ref{eq:I_phi}). Shown from left to right are a cylinder, a truncated sphere and a cone. The most relevant fit parameters are the radii $R_0$ and the tilt angle $\alpha$ of the cone.}
 \label{fig:particle_shapes}
\end{figure*}

In order to obtain information about the island shape, the scattering intensity will be analyzed at a particular $q_z$ but in real space
on the plane. Hence one needs a two--dimensional inverse Fourier transform (${\rm FT}_{2D}^{-1}$) of Eq.~(\ref{eq:Idiff2}).
We introduce
\begin{align}
\centering
\Phi(\vec r_{||},q_z) = {\rm FT}_{2D}^{-1} |F(\vec q_{||}, q_z)|^2 
\label{Eq:phi_ff}
\end{align}
which is related to the (real--space) island shape function $\Omega_s$ by
\begin{align}
\centering
\Phi= ({\rm FT}_z \Omega_s) \otimes ({\rm FT}_z \Omega_s)^\star \;.
\label{eq:phi}
\end{align} 
Here, ${\rm FT}_z$ is the Fourier transform in $z$--direction,
 $\otimes$ denotes the two-dimensional, in--plane convolution and $^\star$ denotes
complex conjugation.
Hence  Eq.~(\ref{eq:Idiff2}) becomes
\begin{align}
\centering
\hat I (\vec r_{||}, q_z) &= & {\rm FT}^{-1}_{2D} I(\vec q_{||},q_z) \qquad \qquad \nonumber \\
                          &\propto& N \left(  \Phi(\vec r_{||},q_z) + \rho \Phi(\vec r_{||},q_z) \otimes h(\vec r_{||})    \right)
\label{eq:IFT}
\end{align}
For our subsequent analysis, we assume an island shape which is isotropic in the plane, hence $\hat I, \Phi, h$ depend only
on $r_{||}=|\vec r_{||}|$.
Determination of the island shape (using the shape function $\Omega_s$ and, derived from it, the function $\Phi$) is easily possible
if the total correlation function in the plane $h$ is sufficiently small and the variations of $h$ occur on larger length scale than those
of $\Phi$. Then one can restrict oneself to regions of small $r_{||}$ where $\hat I$ is dominated by the island form factor contribution
and assume
\begin{align}
 \centering
 \hat{I}(r_{||}, q_z) \propto N\,\Phi( r_{||},q_z)  \qquad ({\rm small}\; r_{||})\,.
 \label{eq:I_phi}
\end{align}
This equation will be used in the following to determine the average island shape. The different island shapes used to model our data are shown in Fig.~\ref{fig:particle_shapes}.

\subsection{Discussion of the GISAXS profiles}
\begin{figure*}[htb]
 \centering
 \includegraphics[width=0.80\textwidth]{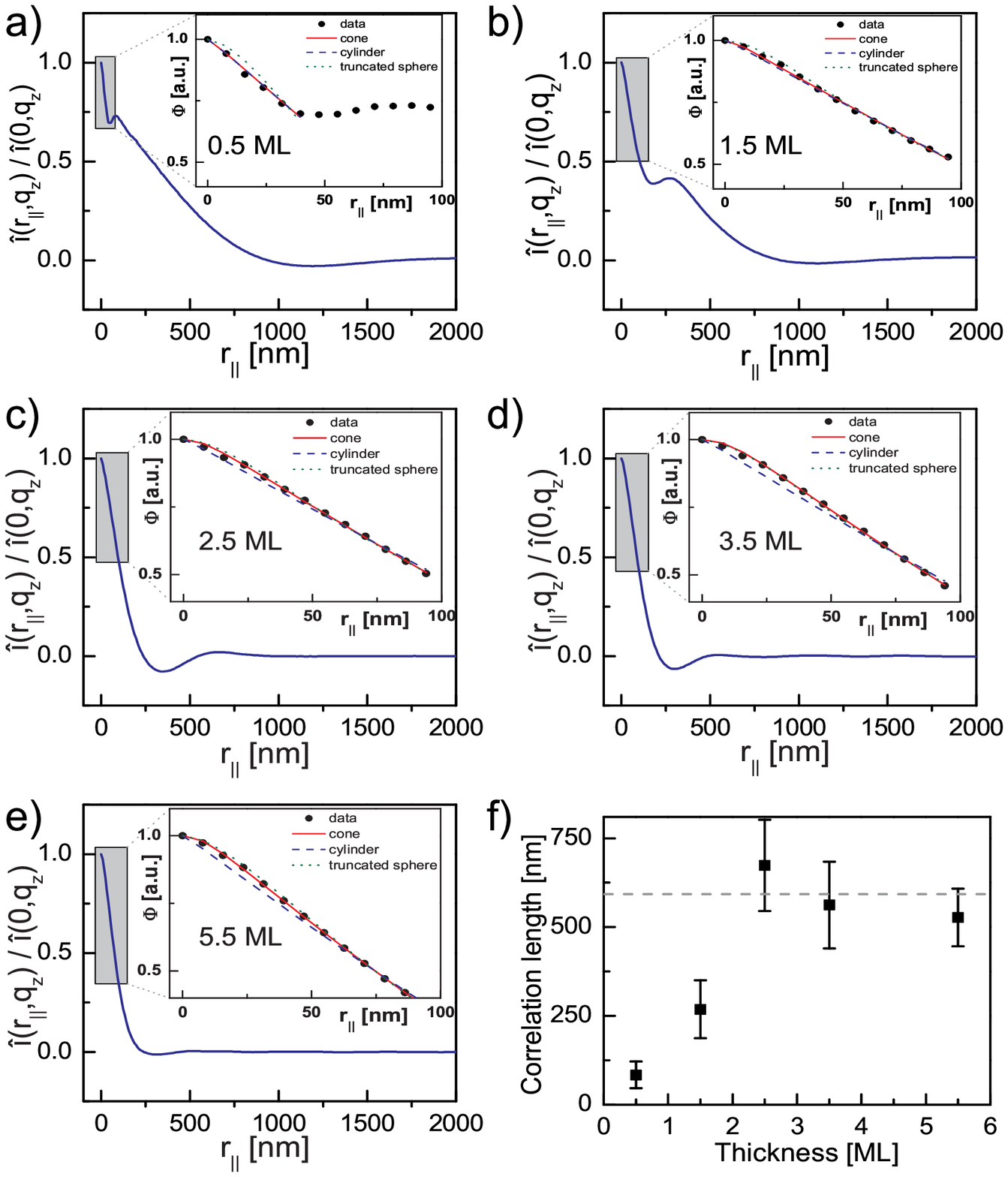}
  \caption{(Color online) (a-e) Normalized Fourier transform (blue solid line) of the line profiles shown in Fig.~\ref{fig:desy_gisaxs} for coverages $\Theta$ of 0.5 to 5.5\,ML (taken at $q_z=0.91\,\rm{nm}^{-1})$. The inset shows a magnification for small radii (indicated by the shaded area) along with the fit of $\Phi$ for the different island shapes. (f) In-plane correlation length observed in the morphology of the samples, as function of film thicknesses. The correlation lengths are obtained from (a-e) and can be related to the average inter-island distance. The dashed line represents the saturation value.}
 \label{fig:DESY_FT}
\end{figure*}

A comprehensive summary of the inverse in--plane  Fourier transform $\hat I(r_{||}, q_z=0.91\,\mathrm{nm}^{-1})$ of the GISAXS line profiles is provided for all samples in Fig.~\ref{fig:DESY_FT}. 
For all coverages 0.5...5.5 ML, $\hat I$ shows a strong variation for small $r_{||}$ ($r_{||} \alt 100$ nm) which are attributable to the island shape.
However, for the lowest coverages of 0.5 and 1.5 ML, $\hat I$ is not small for larger lengths and shows a decay on the scale of several hundred nm.
It is actually difficult to explain this behavior of $\hat I$ using Eq.~(\ref{eq:IFT}) which is based on a model of monodisperse islands having
a size of about 100 nm. Indeed, the AFM images for these low coverages (see Fig.~\ref{fig:desy_afm} (a) and (b)) reveal that islands are
quite dense and are coalesced to greater, anisotropic objects with different sizes. This case should be more appropriately analyzed with a model taking into account
the polydispersity of such fused islands, and we expect that fitting an island shape through Eq.~(\ref{eq:I_phi}) is a first guess.  
For higher coverages (2.5 to 5.5 ML) the separation of form factor and pair correlation contribution is much clearer in the results for $\hat I$.
From the behavior at $r_{||}>200$ nm, one can infer that the total correlation function $h$ (describing the normalized density correlations of the islands) shows small oscillations with a relative strength of about 10\% of the maximum value of $\hat I$ with a first minimum at about 250 nm. Pair correlation functions of differently interacting objects and their oscillatory structure in $r_{||}$ have been intensively studied (see Ref.~\onlinecite{Hansen_2006_book}) and we find that the observed behavior is, in fact, consistent with that of hard objects of about 100 nm size at small to moderate densities.    
The positions of the first maxima of $\hat I$ in Fig.~\ref{fig:DESY_FT}(a-e) account for an in-plane correlation length within the sample surface and can be related to the average (nearest neighbor) island-to-island distance.  

The dependence of the inter-island distance (or inter-island spacing) on the film coverage, as derived from the positions of the first maxima, 
is demonstrated in Fig.~\ref{fig:DESY_FT}(f). We observe that the inter-island spacing 
for small coverages (0.5 and 1.5 ML) is substantially smaller than the inter-island spacing for higher coverages. 
In view of the discussion above, the small distance for 0.5 ML ($\approx 80$ nm) is just the distance of coalesced islands, while for 1.5 ML ($\approx 250$ nm) 
it is related to both coalescence and distance between islands since the islands are quite dense.
In the case of 3D-growth (for small film thicknesses) the inter-island distance does not depend on the amount of deposited material, since mounds only grow higher but do not change their lateral positions. Therefore, the saturation value of $\simeq 600$\,nm, which is achieved for thicknesses $\Theta \geq 2.5$\,ML, points to the formation of 3D-islands. 
These observations are consistent with the \textit{in situ} evolution of the inter-island distance reported in Ref.~[\onlinecite{Frank_2014_dip-realtime}]. 

\subsection{Fitting the GISAXS profiles}
\label{chap:dip_post_morph}
Based on the theory discussed in the previous section, the mean island shape and size can be determined from the functional dependence of $\hat{I}(r_{||}, q_z)={\rm FT}^{-1}_{2D} I(q_{||},q_z)$ close to $r_{||}=0$. Since $\Delta r_{||}$, i.e.\ the minimum resolvable distance between two points in real space, is given by $\pi/q_{||,\mathrm{max}}$, a sufficient experimental resolution and maximum momentum transfer are required to satisfy the Nyquist-Shannon sampling theorem~\cite{Shannon_1948} and to obtain information close to $r_{||}\simeq 0$. 
 
In our model, $\hat I$ close to $r_{||}=0$ is only determined by the island shape function (see Eqs.~(\ref{eq:phi}) and
(\ref{eq:I_phi})). When using hard shapes with rotational symmetry as depicted in Fig.~\ref{fig:particle_shapes}, the necessary
in--plane convolutions in $\hat I$ could also be computed analytically (overlapping circles), and an additional $z$--integration 
has to be done numerically.  
Here, we have chosen to fully numerically compute $\hat I$ by using Fast Fourier transformation techniques in Eq.~(\ref{Eq:phi_ff}). 
Owing to the cylindrical symmetry of the islands, the 2D Fourier transform for the ${||}$-component can actually be related to 
a Hankel transform, when using polar coordinates. 
The Hankel transform is most efficently  
calculated by the logFFT algorithm~\cite{Hamilton_2000,Botan_2000_PRE}, i.e.\ 
by solving the Fast Fourier Transform (FFT) on a logarithmic grid. 
Note that from a numerical perspective, the Hankel transform is ideally suited to analyze the full 
range of $\hat I$, i.e.\ taking the correlations of the islands $h(r)$ into account, or
to analyze ``fuzzy'' shape functions which would arise when one orientationally averages over fused islands. This 
is however beyond this study but potentially interesting for future work.
Accordingly, utilizing the scientific computational software package ``MATLAB''~\cite{MATLAB:2012}, $\Phi$ has been calculated using the form factor of either a cylinder $F_{\mathrm{cyl}}$, a cone $F_{\mathrm{co}}$, or a truncated sphere $F_{\mathrm{sph}}$, given by
\begin{align}
 \centering
 F_{\mathrm{cyl}}&=2\pi R_{0}^2 H_{0} \frac{J_1(q_{||}R_0)}{q_{||}R_0} \mathrm{sinc}(q_z H_0 /2) e^{i q_z H_0 /2}\, , \\
 F_{\mathrm{co}}&=\int_0^{H_0}2\pi R_z^2\frac{J_1(q_{||}R_z)}{q_{||}R_z} e^{i q_z z}dz\, , \\ \notag
 R_z&=R_0-z/\tan{\alpha^\prime},\hspace{0.5cm} H_0/R_0<\tan{\alpha^\prime} \\
 F_{\mathrm{sph}}&=e^{i q_z (H_0 - R_0)}\int_{R_0 - H_0}^H 2\pi R_z^2 \frac{J_1(q_{||}R_z)}{q_{||}R_z}e^{iq_z z}dz\, , \\ \notag
 R_z&=\sqrt{R_0^2-z^2},\hspace{0.5cm} 0<H_0/R_0<2
 \label{eq:I_phi}
\end{align}
where the first order Bessel function $J_1(x)$, the cardinal sinus $\mathrm{sinc}(x)=\sin(x)/x$ and $\alpha^\prime=90^\circ-\alpha$ have been introduced.
A summary of the prevailing theoretical descriptions and a collection of form factors for different island shapes can, for instance be found in Ref.~[\onlinecite{Daillant_2009_book}]. 
To fit the data, the autocorrelation functions for the respective island types were implemented in the MATLAB software package ``Mfit''~\cite{Mfit:2005}.
The optimization of the parameters was then performed using the Simplex Nelder-Mead algorithm~\cite{Nelder_1965}, which is ideally suited to minimize higher dimensional problems.
The resulting deviation from the dataset yields the $\chi^2$-parameter for the goodness of the fit. Based on the AFM-images shown in Fig.~\ref{fig:desy_afm} we have determined the height $H_0$ of the islands for each thickness and subsequently used this as an initial guess for the fitting. As expected, it turned out that $\Phi$ is not very sensitive to the island height within our investigated thickness regime of 0.5--5.5\,ML. Thus, depending on the island type, the base radius $R_0$ and, in the case of the cone also the tilt angle $\alpha$, are the relevant parameters to describe $\Phi$. A description of the parameters and the island shapes used for the fitting is shown in Fig.~\ref{fig:particle_shapes} and a complete summary of the $\chi^2$-parameter for all the fits is given in Tab.~\ref{tab:fiterror}.

\begin{table}
\caption{\label{tab:fiterror}Reduced $\chi^2$-parameters obtained by fitting the data in Fig.~\ref{fig:DESY_FT} with different island shapes. Note that we assumed constant weighting factors for all data points in $\Phi(r_{||},q_z)$.}
\begin{ruledtabular}
\begin{tabular}{l||c|c|c}
Thickness & Cylinder & Tr. Sphere & Cone\\ 
   & Model & Model & Model \\
  $[\mathrm{ML}]$ & [$\chi^2$] & [$\chi^2$] & [$\chi^2$] \\
\hline
   0.5 & 287.9 & 2393.8 & 384.0\\
   1.5 & 120.5 & 159.3 & 83.6\\
   2.5 & 368.3 & 143.8 & 62.2\\
   3.5 & 902.2 & 116.5 & 45.4\\
   5.5 & 781.3 & 148.5 & 25.9\\ 
\end{tabular}
\end{ruledtabular}
\end{table}

\subsubsection{Analysis of the morphology and comparison with AFM}
\begin{figure*}[htb]
 \centering
 \includegraphics[width=0.80\textwidth]{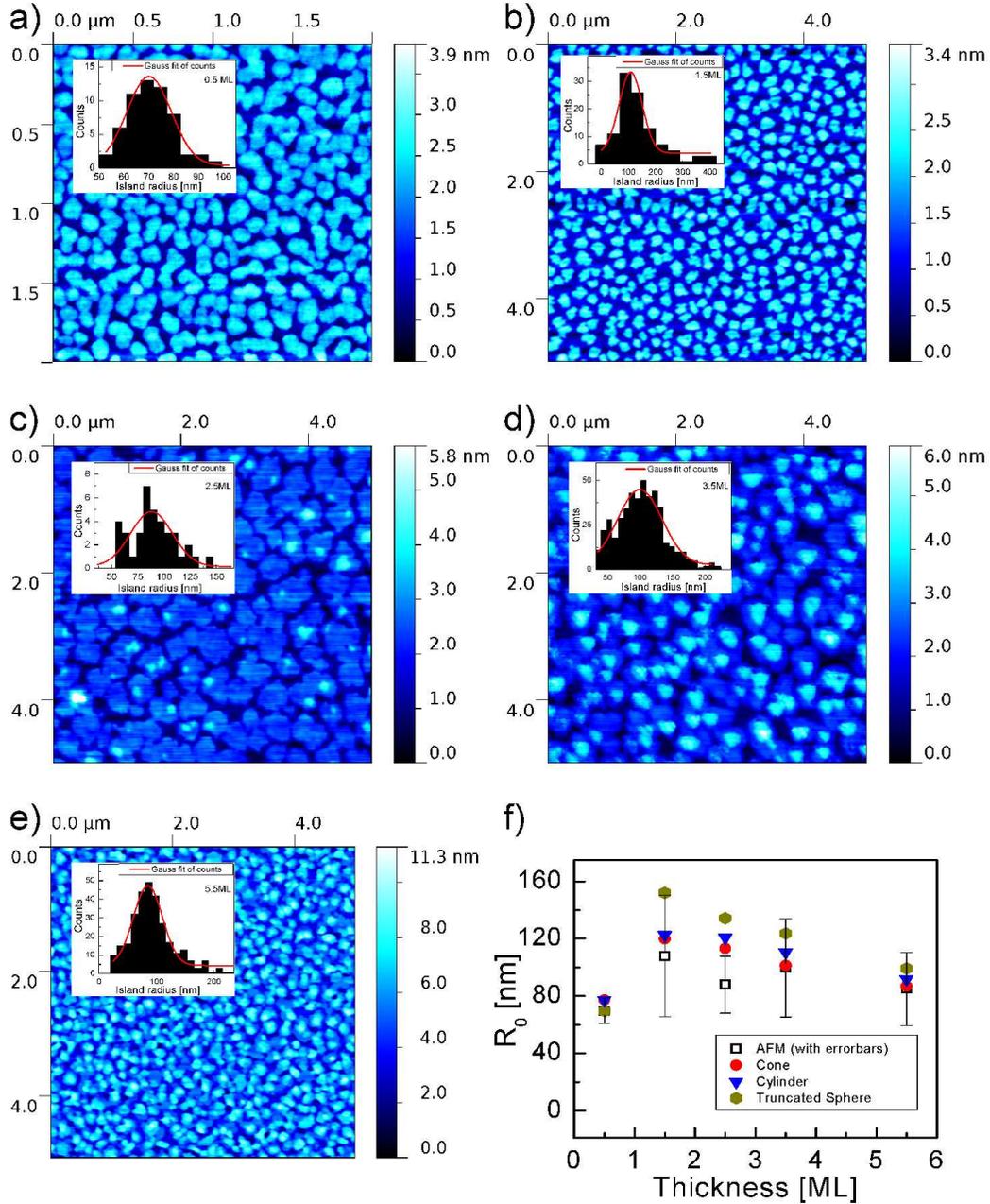}
 \caption{(Color online) (a-e) AFM images for $\Theta=0.5\mbox{--}5.5$\,ML. The inset contains the respective island-size distribution along with a Gaussian fit. The images for 0.5\,ML and 2.5\,ML show a signature of the coalescing of islands, which suggest that the top layer filling is slightly larger than 50\%. (f) Comparison of the thickness dependent radius obtained by AFM (black squares with error bars accounting for the polydispersity) and the GISAXS fits using different island shapes. Among the models used the cone model provides the best description for most of the thicknesses.}
 \label{fig:desy_afm}
\end{figure*}

In order to compare the different island shapes AFM measurements for all \textit{ex situ} samples are shown in Fig.~\ref{fig:desy_afm}. A close inspection of the images taken for $\Theta=0.5$\,ML and $\Theta=2.5$\,ML reveals that islands already coalesce, which suggests that here the top layer coverages are slightly larger than $50$\%. The insets of Fig.~\ref{fig:desy_afm} show the island size distribution for the top structures along with a Gaussian fit to obtain the mean island size. From the full width at half maximum (FWHM) of the Gaussian fits one finds the polydispersity of islands, which is largest for $\Theta=1.5$\,ML.
Remarkably, we observe that the radii obtained by the GISAXS fits (see Fig.~\ref{fig:desy_afm}(f)) match very well with the AFM radii of the structures on top of the islands. This becomes most evident for $\Theta=2.5$\,ML (see Fig.~\ref{fig:desy_afm}(c)) where the bottom terraces are of the order of $\sim 500$\,nm, whereas the top structures are of the order of $\sim 100$\,nm. We suppose that the contrast in the electron density is much higher in the very top layers as compared to the bottom layers. These are practically filled (especially at $\Theta=2.5$\,ML) and act like the ``substrate'' for the mounds. Thus, the main contribution to the scattering signal comes from the top layers allowing the GISAXS technique to primarily sense the top structures.

The insets of Fig.~\ref{fig:DESY_FT} provide a summary of the results obtained by fitting the fall-off of $\hat I(r_{||},q_z)$ normalized to $\hat I\left(r_{||}=0,q_z\right)$ with the function $\Phi(\vec{r}_{||},q_z)$.
From the fits in Fig.~\ref{fig:DESY_FT} (summarized in Tab.~\ref{tab:fiterror}) we observe that the modeling with the cone model provides an excellent agreement at higher film coverages. In contrast, we find that the goodness of the fit, i.e.\ $\chi^2$, becomes worse towards small coverages. For the fits with the cylinder model it is the other way around, i.e.\ we obtain better agreement at low film coverages.

\begin{figure}[!htb]
 \centering
 \includegraphics[width=0.45\textwidth]{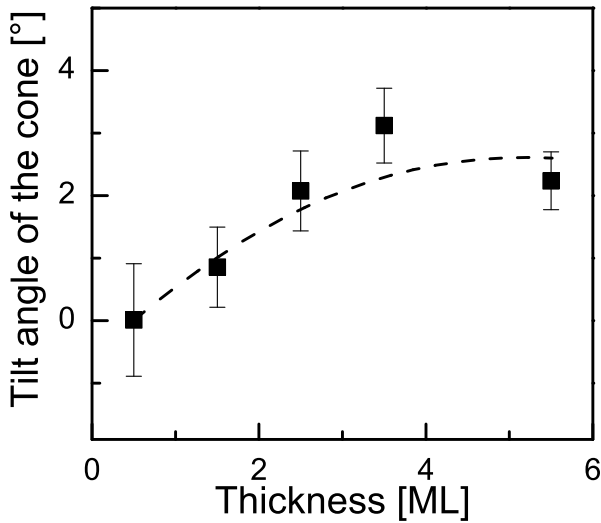}
 \caption{The tilt angle of the cones is shown as a function of the film thickness. Smaller film thicknesses favor a cylinderlike island shape, whereas larger ones deviate from that shape. Note that the dashed line should only serve as a guide to the eye.}
 \label{fig:desy_fit}
\end{figure}

The preceding discussion allows to compare the mean island radii obtained by AFM with those obtained by GISAXS as shown in Fig.~\ref{fig:desy_afm}(f). It may be noted however that the AFM technique can have limitations due to tip convolution effects and inability to probe buried structures. In such cases the GISAXS technique and the analysis provided offers a more conclusive picture. For the cylinder and the cone model we observe very good agreement for all coverages except for $\Theta=2.5$\,ML. However, as mentioned above, this sample particularly features a strong signature of island coalescence and thus a precise statistical determination of the island size is difficult with both methods. We note that the truncated sphere model overestimates the island radius significantly for coverages $\Theta > 0.5$\,ML. For the sample, which shows the worst fit, i.e.\ $\Theta=2.5$\,ML, the cone model deviates by $\sim$22\% from the AFM data, while the truncated sphere model already differs by $\sim$35\%.    
Therefore, we conclude that the truncated sphere model is physically less relevant.
Henceforth, the following discussion will only focus on the cylinder and cone shapes. From our observations these models provide a better parameterization of the film surface.

\subsubsection{Thickness dependence of the island shape}
To further exploit how the morphology varies, we show the tilt angle $\alpha$ (obtained from fitting our data with the cone model) versus the film coverage in Fig.~\ref{fig:desy_fit}. As intuitively expected, we find that for low thicknesses ($\Theta \leq 1.5$\,ML) the tilt angle $\alpha$ is very close to $0\,^{\circ}$ corresponding to a cylinder (i.e.\ a cone with a tilt angle of $0\,^{\circ}$). Evidently, the tilt angle increases only slightly for higher coverages. However, we observe that such a slight increase, already, has a large impact on the shape of the islands: 
obviously, a significant deviation from the cylinder shape is necessary to suit the moundlike surface morphology of the growing islands (see also Tab.~\ref{tab:fiterror}).

This also complies with the results of Ref.~[\onlinecite{Frank_2014_dip-realtime}] where we observe a layer-by-layer growth until a film coverage of 2\,ML and subsequently a transition to a 3D growth. Therefore, we suppose a growth scenario of the following type: at low coverages (i.e.\ $\Theta\leq 2$\,ML) a relatively large interlayer mass transport results in 2D (cylinderlike) islands, which increase in size as the growth progresses and finally merge into non-geometric anisotropic shapes, thus accounting for a layer-by-layer growth. At high coverages (i.e.\ $\Theta > 2$\,ML) the interlayer transport is significantly smaller, which leads to the formation of mounds, therefore we observe a measurable deviation from the cylinder shape. An extreme case would be met, if the interlayer mass transport is completely suppressed resulting then in Poisson-growth. Both, this intermediate case of ``wedding-cake'' structures and the extreme Poisson-growth have previously been observed~\cite{TMichely_2004_book}.

Further improvement of the fits may be obtained e.g.\ by modeling $\Phi$ with two partly merged cylinders in the very low coverage regime. Here, islands show a signature of coalescence and a precise evaluation is more difficult. However, implementing such a complex function is rather challenging and is beyond the scope of the present study.

\section{Conclusion}
This work presents GISAXS measurements of ultra-thin films (i.e.\ below 6\,ML) of the organic semiconductor DIP grown on native SiO$_x$-substrates. Using the shape function of a cone, a cylinder and a truncated sphere we model the DIP-islands and determine the mean radius. This allows for a comparison with the island size distribution obtained by AFM measurements. Remarkably, we observe a significant difference between the employed models. We find that for low film coverages a cylinder provides a suitable shape function to model the islands. Towards higher thicknesses a continuous deviation from the cylinder shape is observed, which is characterized by the formation of wedding-cake structures in the film morphology. 

This work in principle provides a comprehensive picture on the complicated nature of thin film growth and in general its dynamics, which are not yet fully deciphered. 
Importantly, we believe that the presented approach is not only useful to analyze the surface morphology of organic thin film growth in real time, but additionally, may successfully be adopted to other materials potentially featuring a very different shape.

\begin{acknowledgments}
  We gratefully acknowledge the financial support of the DFG. J.N. was supported by the project CEITEC (CZ.1.05/1.1.00/02.0068) from European Regional Development Fund and by the EC 7th Framework Programme (286154/SYLICA). Parts of this research were carried out at the light source PETRA III at DESY, a member of the Helmholtz Association (HGF).  
\end{acknowledgments}


\begin{thebibliography}{64}%
\makeatletter
\providecommand \@ifxundefined [1]{%
 \@ifx{#1\undefined}
}%
\providecommand \@ifnum [1]{%
 \ifnum #1\expandafter \@firstoftwo
 \else \expandafter \@secondoftwo
 \fi
}%
\providecommand \@ifx [1]{%
 \ifx #1\expandafter \@firstoftwo
 \else \expandafter \@secondoftwo
 \fi
}%
\providecommand \natexlab [1]{#1}%
\providecommand \enquote  [1]{``#1''}%
\providecommand \bibnamefont  [1]{#1}%
\providecommand \bibfnamefont [1]{#1}%
\providecommand \citenamefont [1]{#1}%
\providecommand \href@noop [0]{\@secondoftwo}%
\providecommand \href [0]{\begingroup \@sanitize@url \@href}%
\providecommand \@href[1]{\@@startlink{#1}\@@href}%
\providecommand \@@href[1]{\endgroup#1\@@endlink}%
\providecommand \@sanitize@url [0]{\catcode `\\12\catcode `\$12\catcode
  `\&12\catcode `\#12\catcode `\^12\catcode `\_12\catcode `\%12\relax}%
\providecommand \@@startlink[1]{}%
\providecommand \@@endlink[0]{}%
\providecommand \url  [0]{\begingroup\@sanitize@url \@url }%
\providecommand \@url [1]{\endgroup\@href {#1}{\urlprefix }}%
\providecommand \urlprefix  [0]{URL }%
\providecommand \Eprint [0]{\href }%
\providecommand \doibase [0]{http://dx.doi.org/}%
\providecommand \selectlanguage [0]{\@gobble}%
\providecommand \bibinfo  [0]{\@secondoftwo}%
\providecommand \bibfield  [0]{\@secondoftwo}%
\providecommand \translation [1]{[#1]}%
\providecommand \BibitemOpen [0]{}%
\providecommand \bibitemStop [0]{}%
\providecommand \bibitemNoStop [0]{.\EOS\space}%
\providecommand \EOS [0]{\spacefactor3000\relax}%
\providecommand \BibitemShut  [1]{\csname bibitem#1\endcsname}%
\let\auto@bib@innerbib\@empty
\bibitem [{\citenamefont {Michely}\ and\ \citenamefont
  {Krug}(2004)}]{TMichely_2004_book}%
  \BibitemOpen
  \bibfield  {author} {\bibinfo {author} {\bibfnamefont {T.}~\bibnamefont
  {Michely}}\ and\ \bibinfo {author} {\bibfnamefont {J.}~\bibnamefont {Krug}},\
  }\href@noop {} {\emph {\bibinfo {title} {Islands, Mounds, and Atoms. Patterns
  and Processes in Crystal Growth Far from Equilibrium}}}\ (\bibinfo
  {publisher} {Springer},\ \bibinfo {address} {Berlin},\ \bibinfo {year}
  {2004})\BibitemShut {NoStop}%
\bibitem [{\citenamefont {Fendrich}\ and\ \citenamefont
  {Krug}(2007)}]{Fendrich_2007_PhysRevB}%
  \BibitemOpen
  \bibfield  {author} {\bibinfo {author} {\bibfnamefont {M.}~\bibnamefont
  {Fendrich}}\ and\ \bibinfo {author} {\bibfnamefont {J.}~\bibnamefont
  {Krug}},\ }\href@noop {} {\bibfield  {journal} {\bibinfo  {journal} {Phys.
  Rev. B}\ }\textbf {\bibinfo {volume} {76}},\ \bibinfo {eid} {121302}
  (\bibinfo {year} {2007})}\BibitemShut {NoStop}%
\bibitem [{\citenamefont {Sinha}\ \emph {et~al.}(1994)\citenamefont {Sinha},
  \citenamefont {Sanyal}, \citenamefont {Satija}, \citenamefont {Majkrzak},
  \citenamefont {Neumann}, \citenamefont {Homma}, \citenamefont {Szpala},
  \citenamefont {Gibaud},\ and\ \citenamefont {Morkoc}}]{Sinha_1994_PhysicaB}%
  \BibitemOpen
  \bibfield  {author} {\bibinfo {author} {\bibfnamefont {S.~K.}\ \bibnamefont
  {Sinha}}, \bibinfo {author} {\bibfnamefont {M.~K.}\ \bibnamefont {Sanyal}},
  \bibinfo {author} {\bibfnamefont {S.~K.}\ \bibnamefont {Satija}}, \bibinfo
  {author} {\bibfnamefont {C.~F.}\ \bibnamefont {Majkrzak}}, \bibinfo {author}
  {\bibfnamefont {D.~A.}\ \bibnamefont {Neumann}}, \bibinfo {author}
  {\bibfnamefont {H.}~\bibnamefont {Homma}}, \bibinfo {author} {\bibfnamefont
  {S.}~\bibnamefont {Szpala}}, \bibinfo {author} {\bibfnamefont
  {A.}~\bibnamefont {Gibaud}}, \ and\ \bibinfo {author} {\bibfnamefont
  {H.}~\bibnamefont {Morkoc}},\ }\href@noop {} {\bibfield  {journal} {\bibinfo
  {journal} {Physica B}\ }\textbf {\bibinfo {volume} {198}},\ \bibinfo {pages}
  {72} (\bibinfo {year} {1994})}\BibitemShut {NoStop}%
\bibitem [{\citenamefont {Woll}, \citenamefont {Desai},\ and\ \citenamefont
  {Engstrom}(2011)}]{Woll_2011_PhysRevB}%
  \BibitemOpen
  \bibfield  {author} {\bibinfo {author} {\bibfnamefont {A.~R.}\ \bibnamefont
  {Woll}}, \bibinfo {author} {\bibfnamefont {T.~V.}\ \bibnamefont {Desai}}, \
  and\ \bibinfo {author} {\bibfnamefont {J.~R.}\ \bibnamefont {Engstrom}},\
  }\href@noop {} {\bibfield  {journal} {\bibinfo  {journal} {Phys. Rev. B}\
  }\textbf {\bibinfo {volume} {84}},\ \bibinfo {pages} {075479} (\bibinfo
  {year} {2011})}\BibitemShut {NoStop}%
\bibitem [{\citenamefont {Kowarik}\ \emph {et~al.}(2008)\citenamefont
  {Kowarik}, \citenamefont {Gerlach}, \citenamefont {Hinderhofer},
  \citenamefont {Milita}, \citenamefont {Borgatti}, \citenamefont {Zontone},
  \citenamefont {Suzuki}, \citenamefont {Biscarini},\ and\ \citenamefont
  {Schreiber}}]{Kowarik_2008_PhysStatusSolidiRRL}%
  \BibitemOpen
  \bibfield  {author} {\bibinfo {author} {\bibfnamefont {S.}~\bibnamefont
  {Kowarik}}, \bibinfo {author} {\bibfnamefont {A.}~\bibnamefont {Gerlach}},
  \bibinfo {author} {\bibfnamefont {A.}~\bibnamefont {Hinderhofer}}, \bibinfo
  {author} {\bibfnamefont {S.}~\bibnamefont {Milita}}, \bibinfo {author}
  {\bibfnamefont {F.}~\bibnamefont {Borgatti}}, \bibinfo {author}
  {\bibfnamefont {F.}~\bibnamefont {Zontone}}, \bibinfo {author} {\bibfnamefont
  {T.}~\bibnamefont {Suzuki}}, \bibinfo {author} {\bibfnamefont
  {F.}~\bibnamefont {Biscarini}}, \ and\ \bibinfo {author} {\bibfnamefont
  {F.}~\bibnamefont {Schreiber}},\ }\href@noop {} {\bibfield  {journal}
  {\bibinfo  {journal} {Phys. Status Solidi RRL}\ }\textbf {\bibinfo {volume}
  {2}},\ \bibinfo {pages} {120} (\bibinfo {year} {2008})}\BibitemShut {NoStop}%
\bibitem [{\citenamefont {Sinha}\ \emph {et~al.}(1996)\citenamefont {Sinha},
  \citenamefont {Feng}, \citenamefont {Melendres}, \citenamefont {Lee},
  \citenamefont {Russell}, \citenamefont {Satija}, \citenamefont {Sirota},\
  and\ \citenamefont {Sanyal}}]{Sinha_1996_PhysicaA}%
  \BibitemOpen
  \bibfield  {author} {\bibinfo {author} {\bibfnamefont {S.~K.}\ \bibnamefont
  {Sinha}}, \bibinfo {author} {\bibfnamefont {Y.~P.}\ \bibnamefont {Feng}},
  \bibinfo {author} {\bibfnamefont {C.~A.}\ \bibnamefont {Melendres}}, \bibinfo
  {author} {\bibfnamefont {D.~D.}\ \bibnamefont {Lee}}, \bibinfo {author}
  {\bibfnamefont {T.~P.}\ \bibnamefont {Russell}}, \bibinfo {author}
  {\bibfnamefont {S.~K.}\ \bibnamefont {Satija}}, \bibinfo {author}
  {\bibfnamefont {E.~B.}\ \bibnamefont {Sirota}}, \ and\ \bibinfo {author}
  {\bibfnamefont {M.~K.}\ \bibnamefont {Sanyal}},\ }\href@noop {} {\bibfield
  {journal} {\bibinfo  {journal} {Physica A}\ }\textbf {\bibinfo {volume}
  {231}},\ \bibinfo {pages} {99} (\bibinfo {year} {1996})}\BibitemShut
  {NoStop}%
\bibitem [{\citenamefont {Pershan}(2000)}]{Pershan_2000_COLLOIDSURFACEA}%
  \BibitemOpen
  \bibfield  {author} {\bibinfo {author} {\bibfnamefont {P.}~\bibnamefont
  {Pershan}},\ }\href@noop {} {\bibfield  {journal} {\bibinfo  {journal}
  {Colloid Surface A}\ }\textbf {\bibinfo {volume} {171}},\ \bibinfo {pages}
  {149} (\bibinfo {year} {2000})}\BibitemShut {NoStop}%
\bibitem [{\citenamefont {Nickel}\ \emph {et~al.}(2004)\citenamefont {Nickel},
  \citenamefont {Barabash}, \citenamefont {Ruiz}, \citenamefont {Koch},
  \citenamefont {Kahn}, \citenamefont {Feldman}, \citenamefont {Haglund},\ and\
  \citenamefont {Scoles}}]{Nickel_2004_PhysRevB}%
  \BibitemOpen
  \bibfield  {author} {\bibinfo {author} {\bibfnamefont {B.}~\bibnamefont
  {Nickel}}, \bibinfo {author} {\bibfnamefont {R.}~\bibnamefont {Barabash}},
  \bibinfo {author} {\bibfnamefont {R.}~\bibnamefont {Ruiz}}, \bibinfo {author}
  {\bibfnamefont {N.}~\bibnamefont {Koch}}, \bibinfo {author} {\bibfnamefont
  {A.}~\bibnamefont {Kahn}}, \bibinfo {author} {\bibfnamefont {L.~C.}\
  \bibnamefont {Feldman}}, \bibinfo {author} {\bibfnamefont {R.~F.}\
  \bibnamefont {Haglund}}, \ and\ \bibinfo {author} {\bibfnamefont
  {G.}~\bibnamefont {Scoles}},\ }\href@noop {} {\bibfield  {journal} {\bibinfo
  {journal} {Phys. Rev. B}\ }\textbf {\bibinfo {volume} {70}},\ \bibinfo
  {pages} {125401} (\bibinfo {year} {2004})}\BibitemShut {NoStop}%
\bibitem [{\citenamefont {Sinha}\ \emph {et~al.}(1988)\citenamefont {Sinha},
  \citenamefont {Sirota}, \citenamefont {Garoff},\ and\ \citenamefont
  {Stanley}}]{Sinha_1988_PhysRevB}%
  \BibitemOpen
  \bibfield  {author} {\bibinfo {author} {\bibfnamefont {S.~K.}\ \bibnamefont
  {Sinha}}, \bibinfo {author} {\bibfnamefont {E.~B.}\ \bibnamefont {Sirota}},
  \bibinfo {author} {\bibfnamefont {S.}~\bibnamefont {Garoff}}, \ and\ \bibinfo
  {author} {\bibfnamefont {H.~B.}\ \bibnamefont {Stanley}},\ }\href@noop {}
  {\bibfield  {journal} {\bibinfo  {journal} {Phys. Rev. B}\ }\textbf {\bibinfo
  {volume} {38}},\ \bibinfo {pages} {2297} (\bibinfo {year}
  {1988})}\BibitemShut {NoStop}%
\bibitem [{\citenamefont {Tolan}(1999)}]{Tolan}%
  \BibitemOpen
  \bibfield  {author} {\bibinfo {author} {\bibfnamefont {M.}~\bibnamefont
  {Tolan}},\ }\href@noop {} {\emph {\bibinfo {title} {X-ray scattering from
  soft-matter thin films: materials science and basic research}}},\ Springer
  tracts in modern physics\ (\bibinfo  {publisher} {Springer},\ \bibinfo
  {address} {Berlin ; London},\ \bibinfo {year} {1999})\BibitemShut {NoStop}%
\bibitem [{\citenamefont {Salditt}, \citenamefont {Metzger},\ and\
  \citenamefont {Peisl}(1994)}]{Salditt_1994_PhysRevLett}%
  \BibitemOpen
  \bibfield  {author} {\bibinfo {author} {\bibfnamefont {T.}~\bibnamefont
  {Salditt}}, \bibinfo {author} {\bibfnamefont {T.}~\bibnamefont {Metzger}}, \
  and\ \bibinfo {author} {\bibfnamefont {J.}~\bibnamefont {Peisl}},\
  }\href@noop {} {\bibfield  {journal} {\bibinfo  {journal} {Phys. Rev. Lett.}\
  }\textbf {\bibinfo {volume} {73}},\ \bibinfo {pages} {2228} (\bibinfo {year}
  {1994})}\BibitemShut {NoStop}%
\bibitem [{\citenamefont {Fleet}\ \emph {et~al.}(2006)\citenamefont {Fleet},
  \citenamefont {Dale}, \citenamefont {Woll}, \citenamefont {Suzuki},\ and\
  \citenamefont {Brock}}]{Fleet_2006_PhysRevLett}%
  \BibitemOpen
  \bibfield  {author} {\bibinfo {author} {\bibfnamefont {A.}~\bibnamefont
  {Fleet}}, \bibinfo {author} {\bibfnamefont {D.}~\bibnamefont {Dale}},
  \bibinfo {author} {\bibfnamefont {A.~R.}\ \bibnamefont {Woll}}, \bibinfo
  {author} {\bibfnamefont {Y.}~\bibnamefont {Suzuki}}, \ and\ \bibinfo {author}
  {\bibfnamefont {J.~D.}\ \bibnamefont {Brock}},\ }\href@noop {} {\bibfield
  {journal} {\bibinfo  {journal} {Phys. Rev. Lett.}\ }\textbf {\bibinfo
  {volume} {96}},\ \bibinfo {pages} {055508} (\bibinfo {year}
  {2006})}\BibitemShut {NoStop}%
\bibitem [{\citenamefont {Frank}\ \emph {et~al.}(2014)\citenamefont {Frank},
  \citenamefont {Nov{\'a}k}, \citenamefont {Banerjee}, \citenamefont {Gerlach},
  \citenamefont {Schreiber}, \citenamefont {Vorobiev},\ and\ \citenamefont
  {Kowarik}}]{Frank_2014_dip-realtime}%
  \BibitemOpen
  \bibfield  {author} {\bibinfo {author} {\bibfnamefont {C.}~\bibnamefont
  {Frank}}, \bibinfo {author} {\bibfnamefont {J.}~\bibnamefont {Nov{\'a}k}},
  \bibinfo {author} {\bibfnamefont {R.}~\bibnamefont {Banerjee}}, \bibinfo
  {author} {\bibfnamefont {A.}~\bibnamefont {Gerlach}}, \bibinfo {author}
  {\bibfnamefont {F.}~\bibnamefont {Schreiber}}, \bibinfo {author}
  {\bibfnamefont {A.}~\bibnamefont {Vorobiev}}, \ and\ \bibinfo {author}
  {\bibfnamefont {S.}~\bibnamefont {Kowarik}},\ }\href@noop {} {\bibfield
  {journal} {\bibinfo  {journal} {Phys. Rev. B}\ }\textbf {\bibinfo {volume}
  {90}},\ \bibinfo {pages} {045410} (\bibinfo {year} {2014})}\BibitemShut
  {NoStop}%
\bibitem [{\citenamefont {Yu}\ \emph {et~al.}(2013)\citenamefont {Yu},
  \citenamefont {Santoro}, \citenamefont {Sarkar}, \citenamefont {Dicke},
  \citenamefont {Wessels}, \citenamefont {Bommel}, \citenamefont
  {D\"{o}hrmann}, \citenamefont {Perlich}, \citenamefont {Kuhlmann},
  \citenamefont {Metwalli}, \citenamefont {Risch}, \citenamefont
  {Schwartzkopf}, \citenamefont {Drescher}, \citenamefont
  {M\"{u}ller-Buschbaum},\ and\ \citenamefont {Roth}}]{Yu_JPhysChemLett_2013}%
  \BibitemOpen
  \bibfield  {author} {\bibinfo {author} {\bibfnamefont {S.}~\bibnamefont
  {Yu}}, \bibinfo {author} {\bibfnamefont {G.}~\bibnamefont {Santoro}},
  \bibinfo {author} {\bibfnamefont {K.}~\bibnamefont {Sarkar}}, \bibinfo
  {author} {\bibfnamefont {B.}~\bibnamefont {Dicke}}, \bibinfo {author}
  {\bibfnamefont {P.}~\bibnamefont {Wessels}}, \bibinfo {author} {\bibfnamefont
  {S.}~\bibnamefont {Bommel}}, \bibinfo {author} {\bibfnamefont
  {R.}~\bibnamefont {D\"{o}hrmann}}, \bibinfo {author} {\bibfnamefont
  {J.}~\bibnamefont {Perlich}}, \bibinfo {author} {\bibfnamefont
  {M.}~\bibnamefont {Kuhlmann}}, \bibinfo {author} {\bibfnamefont
  {E.}~\bibnamefont {Metwalli}}, \bibinfo {author} {\bibfnamefont {J.~F.~H.}\
  \bibnamefont {Risch}}, \bibinfo {author} {\bibfnamefont {M.}~\bibnamefont
  {Schwartzkopf}}, \bibinfo {author} {\bibfnamefont {M.}~\bibnamefont
  {Drescher}}, \bibinfo {author} {\bibfnamefont {P.}~\bibnamefont
  {M\"{u}ller-Buschbaum}}, \ and\ \bibinfo {author} {\bibfnamefont {S.~V.}\
  \bibnamefont {Roth}},\ }\href@noop {} {\bibfield  {journal} {\bibinfo
  {journal} {J. Phys. Chem. Lett.}\ }\textbf {\bibinfo {volume} {4}},\ \bibinfo
  {pages} {3170} (\bibinfo {year} {2013})}\BibitemShut {NoStop}%
\bibitem [{\citenamefont {Santoro}\ \emph
  {et~al.}(2014{\natexlab{a}})\citenamefont {Santoro}, \citenamefont {Yu},
  \citenamefont {Schwartzkopf}, \citenamefont {Zhang}, \citenamefont {Vayalil},
  \citenamefont {Risch}, \citenamefont {R\"{u}bhausen}, \citenamefont
  {Hern\'{a}ndez}, \citenamefont {Domingo},\ and\ \citenamefont
  {Roth}}]{Gonzalo_APL_2014}%
  \BibitemOpen
  \bibfield  {author} {\bibinfo {author} {\bibfnamefont {G.}~\bibnamefont
  {Santoro}}, \bibinfo {author} {\bibfnamefont {S.}~\bibnamefont {Yu}},
  \bibinfo {author} {\bibfnamefont {M.}~\bibnamefont {Schwartzkopf}}, \bibinfo
  {author} {\bibfnamefont {P.}~\bibnamefont {Zhang}}, \bibinfo {author}
  {\bibfnamefont {S.~K.}\ \bibnamefont {Vayalil}}, \bibinfo {author}
  {\bibfnamefont {J.~F.~H.}\ \bibnamefont {Risch}}, \bibinfo {author}
  {\bibfnamefont {M.~A.}\ \bibnamefont {R\"{u}bhausen}}, \bibinfo {author}
  {\bibfnamefont {M.}~\bibnamefont {Hern\'{a}ndez}}, \bibinfo {author}
  {\bibfnamefont {C.}~\bibnamefont {Domingo}}, \ and\ \bibinfo {author}
  {\bibfnamefont {S.~V.}\ \bibnamefont {Roth}},\ }\href@noop {} {\bibfield
  {journal} {\bibinfo  {journal} {Appl. Phys. Lett.}\ }\textbf {\bibinfo
  {volume} {104}},\ \bibinfo {pages} {243107} (\bibinfo {year}
  {2014}{\natexlab{a}})}\BibitemShut {NoStop}%
\bibitem [{\citenamefont {Frank}\ \emph {et~al.}(2013)\citenamefont {Frank},
  \citenamefont {Nov{\'a}k}, \citenamefont {Gerlach}, \citenamefont {Ligorio},
  \citenamefont {Broch}, \citenamefont {Hinderhofer}, \citenamefont
  {Aufderheide}, \citenamefont {Banerjee}, \citenamefont {Nervo},\ and\
  \citenamefont {Schreiber}}]{Frank_2013_JApplPhys}%
  \BibitemOpen
  \bibfield  {author} {\bibinfo {author} {\bibfnamefont {C.}~\bibnamefont
  {Frank}}, \bibinfo {author} {\bibfnamefont {J.}~\bibnamefont {Nov{\'a}k}},
  \bibinfo {author} {\bibfnamefont {A.}~\bibnamefont {Gerlach}}, \bibinfo
  {author} {\bibfnamefont {G.}~\bibnamefont {Ligorio}}, \bibinfo {author}
  {\bibfnamefont {K.}~\bibnamefont {Broch}}, \bibinfo {author} {\bibfnamefont
  {A.}~\bibnamefont {Hinderhofer}}, \bibinfo {author} {\bibfnamefont
  {A.}~\bibnamefont {Aufderheide}}, \bibinfo {author} {\bibfnamefont
  {R.}~\bibnamefont {Banerjee}}, \bibinfo {author} {\bibfnamefont
  {R.}~\bibnamefont {Nervo}}, \ and\ \bibinfo {author} {\bibfnamefont
  {F.}~\bibnamefont {Schreiber}},\ }\href@noop {} {\bibfield  {journal}
  {\bibinfo  {journal} {J. Appl. Phys.}\ }\textbf {\bibinfo {volume} {114}},\
  \bibinfo {pages} {043515} (\bibinfo {year} {2013})}\BibitemShut {NoStop}%
\bibitem [{\citenamefont {Lazzari}, \citenamefont {Leroy},\ and\ \citenamefont
  {Renaud}(2007)}]{Lazzari_2007_PhysRevB}%
  \BibitemOpen
  \bibfield  {author} {\bibinfo {author} {\bibfnamefont {R.}~\bibnamefont
  {Lazzari}}, \bibinfo {author} {\bibfnamefont {F.}~\bibnamefont {Leroy}}, \
  and\ \bibinfo {author} {\bibfnamefont {G.}~\bibnamefont {Renaud}},\
  }\href@noop {} {\bibfield  {journal} {\bibinfo  {journal} {Phys. Rev. B}\
  }\textbf {\bibinfo {volume} {76}},\ \bibinfo {pages} {125411} (\bibinfo
  {year} {2007})}\BibitemShut {NoStop}%
\bibitem [{\citenamefont {M\"uller-Buschbaum}(2009)}]{Buschbaum_2009_Book}%
  \BibitemOpen
  \bibfield  {author} {\bibinfo {author} {\bibfnamefont {P.}~\bibnamefont
  {M\"uller-Buschbaum}},\ }in\ \href@noop {} {\emph {\bibinfo {booktitle}
  {Applications of Synchrotron Light to Scattering and Diffraction in Materials
  and Life Sciences}}},\ \bibinfo {series} {Lecture Notes in Physics}, Vol.\
  \bibinfo {volume} {776},\ \bibinfo {editor} {edited by\ \bibinfo {editor}
  {\bibfnamefont {M.~A.}\ \bibnamefont {Gomez}}, \bibinfo {editor}
  {\bibfnamefont {A.}~\bibnamefont {Nogales}}, \bibinfo {editor} {\bibfnamefont
  {M.~C.}\ \bibnamefont {Garcia-Gutierrez}}, \ and\ \bibinfo {editor}
  {\bibfnamefont {T.}~\bibnamefont {Ezquerra}}}\ (\bibinfo  {publisher}
  {Springer},\ \bibinfo {address} {Berlin Heidelberg},\ \bibinfo {year}
  {2009})\ pp.\ \bibinfo {pages} {61--89}\BibitemShut {NoStop}%
\bibitem [{\citenamefont {Smilgies}\ \emph {et~al.}(2002)\citenamefont
  {Smilgies}, \citenamefont {Busch}, \citenamefont {Papadakis},\ and\
  \citenamefont {Posselt}}]{Smilgies_2002_SynchrotronRadiationNews}%
  \BibitemOpen
  \bibfield  {author} {\bibinfo {author} {\bibfnamefont {D.-M.}\ \bibnamefont
  {Smilgies}}, \bibinfo {author} {\bibfnamefont {P.}~\bibnamefont {Busch}},
  \bibinfo {author} {\bibfnamefont {C.~M.}\ \bibnamefont {Papadakis}}, \ and\
  \bibinfo {author} {\bibfnamefont {D.}~\bibnamefont {Posselt}},\ }\href@noop
  {} {\bibfield  {journal} {\bibinfo  {journal} {Synchrotron Radiation News}\
  }\textbf {\bibinfo {volume} {15}},\ \bibinfo {pages} {35} (\bibinfo {year}
  {2002})}\BibitemShut {NoStop}%
\bibitem [{\citenamefont {Metzger}\ \emph {et~al.}(1999)\citenamefont
  {Metzger}, \citenamefont {Kegel}, \citenamefont {Paniago},\ and\
  \citenamefont {Peisl}}]{Metzger_1999_JPhysDApplPhys}%
  \BibitemOpen
  \bibfield  {author} {\bibinfo {author} {\bibfnamefont {T.~H.}\ \bibnamefont
  {Metzger}}, \bibinfo {author} {\bibfnamefont {I.}~\bibnamefont {Kegel}},
  \bibinfo {author} {\bibfnamefont {R.}~\bibnamefont {Paniago}}, \ and\
  \bibinfo {author} {\bibfnamefont {J.}~\bibnamefont {Peisl}},\ }\href@noop {}
  {\bibfield  {journal} {\bibinfo  {journal} {J. Phys. D Appl. Phys.}\ }\textbf
  {\bibinfo {volume} {32}},\ \bibinfo {pages} {A202} (\bibinfo {year}
  {1999})}\BibitemShut {NoStop}%
\bibitem [{\citenamefont {Du}\ \emph {et~al.}(2004)\citenamefont {Du},
  \citenamefont {Li}, \citenamefont {Douki}, \citenamefont {Li}, \citenamefont
  {Garcia}, \citenamefont {Jain}, \citenamefont {Smilgies}, \citenamefont
  {Fetters}, \citenamefont {Gruner}, \citenamefont {Wiesner},\ and\
  \citenamefont {Ober}}]{Du_2004_AdvMater}%
  \BibitemOpen
  \bibfield  {author} {\bibinfo {author} {\bibfnamefont {P.}~\bibnamefont
  {Du}}, \bibinfo {author} {\bibfnamefont {M.}~\bibnamefont {Li}}, \bibinfo
  {author} {\bibfnamefont {K.}~\bibnamefont {Douki}}, \bibinfo {author}
  {\bibfnamefont {X.}~\bibnamefont {Li}}, \bibinfo {author} {\bibfnamefont
  {C.~B.~W.}\ \bibnamefont {Garcia}}, \bibinfo {author} {\bibfnamefont
  {A.}~\bibnamefont {Jain}}, \bibinfo {author} {\bibfnamefont {D.-M.}\
  \bibnamefont {Smilgies}}, \bibinfo {author} {\bibfnamefont {L.~J.}\
  \bibnamefont {Fetters}}, \bibinfo {author} {\bibfnamefont {S.~M.}\
  \bibnamefont {Gruner}}, \bibinfo {author} {\bibfnamefont {U.}~\bibnamefont
  {Wiesner}}, \ and\ \bibinfo {author} {\bibfnamefont {C.~K.}\ \bibnamefont
  {Ober}},\ }\href@noop {} {\bibfield  {journal} {\bibinfo  {journal} {Adv.
  Mater.}\ }\textbf {\bibinfo {volume} {16}},\ \bibinfo {pages} {953} (\bibinfo
  {year} {2004})}\BibitemShut {NoStop}%
\bibitem [{\citenamefont {Stangl}\ \emph {et~al.}(1999)\citenamefont {Stangl},
  \citenamefont {Hol\'{y}}, \citenamefont {Mikul\'{i}k}, \citenamefont {Bauer},
  \citenamefont {Kegel}, \citenamefont {Metzger}, \citenamefont {Schmidt},
  \citenamefont {Lange},\ and\ \citenamefont {Eberl}}]{Stangl_1999_APL}%
  \BibitemOpen
  \bibfield  {author} {\bibinfo {author} {\bibfnamefont {J.}~\bibnamefont
  {Stangl}}, \bibinfo {author} {\bibfnamefont {V.}~\bibnamefont {Hol\'{y}}},
  \bibinfo {author} {\bibfnamefont {P.}~\bibnamefont {Mikul\'{i}k}}, \bibinfo
  {author} {\bibfnamefont {G.}~\bibnamefont {Bauer}}, \bibinfo {author}
  {\bibfnamefont {I.}~\bibnamefont {Kegel}}, \bibinfo {author} {\bibfnamefont
  {T.~H.}\ \bibnamefont {Metzger}}, \bibinfo {author} {\bibfnamefont {O.~G.}\
  \bibnamefont {Schmidt}}, \bibinfo {author} {\bibfnamefont {C.}~\bibnamefont
  {Lange}}, \ and\ \bibinfo {author} {\bibfnamefont {K.}~\bibnamefont
  {Eberl}},\ }\href@noop {} {\bibfield  {journal} {\bibinfo  {journal} {Appl.
  Phys. Lett.}\ }\textbf {\bibinfo {volume} {74}},\ \bibinfo {pages} {3785}
  (\bibinfo {year} {1999})}\BibitemShut {NoStop}%
\bibitem [{\citenamefont {Kurrle}\ and\ \citenamefont
  {Pflaum}(2008)}]{Kurrle_2008_APL}%
  \BibitemOpen
  \bibfield  {author} {\bibinfo {author} {\bibfnamefont {D.}~\bibnamefont
  {Kurrle}}\ and\ \bibinfo {author} {\bibfnamefont {J.}~\bibnamefont
  {Pflaum}},\ }\href@noop {} {\bibfield  {journal} {\bibinfo  {journal} {Appl.
  Phys. Lett.}\ }\textbf {\bibinfo {volume} {92}},\ \bibinfo {pages} {133306}
  (\bibinfo {year} {2008})}\BibitemShut {NoStop}%
\bibitem [{\citenamefont {de~Oteyza}\ \emph {et~al.}(2009)\citenamefont
  {de~Oteyza}, \citenamefont {Garc\'{i}a-Lastra}, \citenamefont {Corso},
  \citenamefont {Doyle}, \citenamefont {Floreano}, \citenamefont {Morgante},
  \citenamefont {Wakayama}, \citenamefont {Rubio},\ and\ \citenamefont
  {Ortega}}]{deOteyza_2009_afm}%
  \BibitemOpen
  \bibfield  {author} {\bibinfo {author} {\bibfnamefont {D.~G.}\ \bibnamefont
  {de~Oteyza}}, \bibinfo {author} {\bibfnamefont {J.~M.}\ \bibnamefont
  {Garc\'{i}a-Lastra}}, \bibinfo {author} {\bibfnamefont {M.}~\bibnamefont
  {Corso}}, \bibinfo {author} {\bibfnamefont {B.~P.}\ \bibnamefont {Doyle}},
  \bibinfo {author} {\bibfnamefont {L.}~\bibnamefont {Floreano}}, \bibinfo
  {author} {\bibfnamefont {A.}~\bibnamefont {Morgante}}, \bibinfo {author}
  {\bibfnamefont {Y.}~\bibnamefont {Wakayama}}, \bibinfo {author}
  {\bibfnamefont {A.}~\bibnamefont {Rubio}}, \ and\ \bibinfo {author}
  {\bibfnamefont {J.~E.}\ \bibnamefont {Ortega}},\ }\href@noop {} {\bibfield
  {journal} {\bibinfo  {journal} {Adv. Funct. Mater.}\ }\textbf {\bibinfo
  {volume} {19}},\ \bibinfo {pages} {3567} (\bibinfo {year}
  {2009})}\BibitemShut {NoStop}%
\bibitem [{\citenamefont {de~Oteyza}\ \emph {et~al.}(2007)\citenamefont
  {de~Oteyza}, \citenamefont {Krauss}, \citenamefont {Barrena}, \citenamefont
  {Sellner}, \citenamefont {Dosch},\ and\ \citenamefont
  {Oss\'{o}}}]{deOteyza_2007_apl}%
  \BibitemOpen
  \bibfield  {author} {\bibinfo {author} {\bibfnamefont {D.~G.}\ \bibnamefont
  {de~Oteyza}}, \bibinfo {author} {\bibfnamefont {T.~N.}\ \bibnamefont
  {Krauss}}, \bibinfo {author} {\bibfnamefont {E.}~\bibnamefont {Barrena}},
  \bibinfo {author} {\bibfnamefont {S.}~\bibnamefont {Sellner}}, \bibinfo
  {author} {\bibfnamefont {H.}~\bibnamefont {Dosch}}, \ and\ \bibinfo {author}
  {\bibfnamefont {J.~O.}\ \bibnamefont {Oss\'{o}}},\ }\href@noop {} {\bibfield
  {journal} {\bibinfo  {journal} {Appl. Phys. Lett.}\ }\textbf {\bibinfo
  {volume} {90}},\ \bibinfo {pages} {243104} (\bibinfo {year}
  {2007})}\BibitemShut {NoStop}%
\bibitem [{\citenamefont {Wagner}\ \emph {et~al.}(2010)\citenamefont {Wagner},
  \citenamefont {Gruber}, \citenamefont {Hinderhofer}, \citenamefont {Wilke},
  \citenamefont {Br\"{o}ker}, \citenamefont {Frisch}, \citenamefont {Amsalem},
  \citenamefont {Vollmer}, \citenamefont {Opitz}, \citenamefont {Koch},
  \citenamefont {Schreiber},\ and\ \citenamefont {Br\"{u}tting}}]{WagnerAM10}%
  \BibitemOpen
  \bibfield  {author} {\bibinfo {author} {\bibfnamefont {J.}~\bibnamefont
  {Wagner}}, \bibinfo {author} {\bibfnamefont {M.}~\bibnamefont {Gruber}},
  \bibinfo {author} {\bibfnamefont {A.}~\bibnamefont {Hinderhofer}}, \bibinfo
  {author} {\bibfnamefont {A.}~\bibnamefont {Wilke}}, \bibinfo {author}
  {\bibfnamefont {B.}~\bibnamefont {Br\"{o}ker}}, \bibinfo {author}
  {\bibfnamefont {J.}~\bibnamefont {Frisch}}, \bibinfo {author} {\bibfnamefont
  {P.}~\bibnamefont {Amsalem}}, \bibinfo {author} {\bibfnamefont
  {A.}~\bibnamefont {Vollmer}}, \bibinfo {author} {\bibfnamefont
  {A.}~\bibnamefont {Opitz}}, \bibinfo {author} {\bibfnamefont
  {N.}~\bibnamefont {Koch}}, \bibinfo {author} {\bibfnamefont {F.}~\bibnamefont
  {Schreiber}}, \ and\ \bibinfo {author} {\bibfnamefont {W.}~\bibnamefont
  {Br\"{u}tting}},\ }\href@noop {} {\bibfield  {journal} {\bibinfo  {journal}
  {Adv. Funct. Mater.}\ }\textbf {\bibinfo {volume} {20}},\ \bibinfo {pages}
  {4295} (\bibinfo {year} {2010})}\BibitemShut {NoStop}%
\bibitem [{\citenamefont {Br\"utting}\ and\ \citenamefont
  {Adachi}(2012)}]{Brutting_2012_book}%
  \BibitemOpen
  \bibfield  {author} {\bibinfo {author} {\bibfnamefont {W.}~\bibnamefont
  {Br\"utting}}\ and\ \bibinfo {author} {\bibfnamefont {C.}~\bibnamefont
  {Adachi}},\ }\href@noop {} {\emph {\bibinfo {title} {{Physics of Organic
  Semiconductors}}}},\ \bibinfo {edition} {2nd}\ ed.\ (\bibinfo  {publisher}
  {Wiley-VCH, Weinheim},\ \bibinfo {year} {2012})\BibitemShut {NoStop}%
\bibitem [{\citenamefont {Tripathi}\ and\ \citenamefont
  {Pflaum}(2006)}]{Tripathi_2006_ApplPhysLett}%
  \BibitemOpen
  \bibfield  {author} {\bibinfo {author} {\bibfnamefont {A.~K.}\ \bibnamefont
  {Tripathi}}\ and\ \bibinfo {author} {\bibfnamefont {J.}~\bibnamefont
  {Pflaum}},\ }\href {\doibase 10.1063/1.2338587} {\bibfield  {journal}
  {\bibinfo  {journal} {Appl. Phys. Lett.}\ }\textbf {\bibinfo {volume} {89}},\
  \bibinfo {pages} {082103} (\bibinfo {year} {2006})}\BibitemShut {NoStop}%
\bibitem [{\citenamefont {Opitz}\ \emph {et~al.}(2010)\citenamefont {Opitz},
  \citenamefont {Wagner}, \citenamefont {Br\"utting}, \citenamefont {Salzmann},
  \citenamefont {Koch}, \citenamefont {Manara}, \citenamefont {Pflaum},
  \citenamefont {Hinderhofer},\ and\ \citenamefont {Schreiber}}]{OpitzIEEE}%
  \BibitemOpen
  \bibfield  {author} {\bibinfo {author} {\bibfnamefont {A.}~\bibnamefont
  {Opitz}}, \bibinfo {author} {\bibfnamefont {J.}~\bibnamefont {Wagner}},
  \bibinfo {author} {\bibfnamefont {W.}~\bibnamefont {Br\"utting}}, \bibinfo
  {author} {\bibfnamefont {I.}~\bibnamefont {Salzmann}}, \bibinfo {author}
  {\bibfnamefont {N.}~\bibnamefont {Koch}}, \bibinfo {author} {\bibfnamefont
  {J.}~\bibnamefont {Manara}}, \bibinfo {author} {\bibfnamefont
  {J.}~\bibnamefont {Pflaum}}, \bibinfo {author} {\bibfnamefont
  {A.}~\bibnamefont {Hinderhofer}}, \ and\ \bibinfo {author} {\bibfnamefont
  {F.}~\bibnamefont {Schreiber}},\ }\href@noop {} {\bibfield  {journal}
  {\bibinfo  {journal} {IEEE J. Sel. Top. Quantum Electron.}\ }\textbf
  {\bibinfo {volume} {16}},\ \bibinfo {pages} {1707} (\bibinfo {year}
  {2010})}\BibitemShut {NoStop}%
\bibitem [{\citenamefont {Hinderhofer}\ \emph {et~al.}(2012)\citenamefont
  {Hinderhofer}, \citenamefont {Hosokai}, \citenamefont {Yonezawa},
  \citenamefont {Gerlach}, \citenamefont {Kato}, \citenamefont {Broch},
  \citenamefont {Frank}, \citenamefont {Nov\'ak}, \citenamefont {Kera},
  \citenamefont {Ueno},\ and\ \citenamefont
  {Schreiber}}]{Hinderhofer_2012_ApplPhysLett}%
  \BibitemOpen
  \bibfield  {author} {\bibinfo {author} {\bibfnamefont {A.}~\bibnamefont
  {Hinderhofer}}, \bibinfo {author} {\bibfnamefont {T.}~\bibnamefont
  {Hosokai}}, \bibinfo {author} {\bibfnamefont {K.}~\bibnamefont {Yonezawa}},
  \bibinfo {author} {\bibfnamefont {A.}~\bibnamefont {Gerlach}}, \bibinfo
  {author} {\bibfnamefont {K.}~\bibnamefont {Kato}}, \bibinfo {author}
  {\bibfnamefont {K.}~\bibnamefont {Broch}}, \bibinfo {author} {\bibfnamefont
  {C.}~\bibnamefont {Frank}}, \bibinfo {author} {\bibfnamefont
  {J.}~\bibnamefont {Nov\'ak}}, \bibinfo {author} {\bibfnamefont
  {S.}~\bibnamefont {Kera}}, \bibinfo {author} {\bibfnamefont {N.}~\bibnamefont
  {Ueno}}, \ and\ \bibinfo {author} {\bibfnamefont {F.}~\bibnamefont
  {Schreiber}},\ }\href@noop {} {\bibfield  {journal} {\bibinfo  {journal}
  {Appl. Phys. Lett.}\ }\textbf {\bibinfo {volume} {101}},\ \bibinfo {pages}
  {033307} (\bibinfo {year} {2012})}\BibitemShut {NoStop}%
\bibitem [{\citenamefont {Kowarik}\ \emph {et~al.}(2009)\citenamefont
  {Kowarik}, \citenamefont {Gerlach}, \citenamefont {Sellner}, \citenamefont
  {Calvacanti},\ and\ \citenamefont {Schreiber}}]{Kowarik_2009_AEM}%
  \BibitemOpen
  \bibfield  {author} {\bibinfo {author} {\bibfnamefont {S.}~\bibnamefont
  {Kowarik}}, \bibinfo {author} {\bibfnamefont {A.}~\bibnamefont {Gerlach}},
  \bibinfo {author} {\bibfnamefont {S.}~\bibnamefont {Sellner}}, \bibinfo
  {author} {\bibfnamefont {L.}~\bibnamefont {Calvacanti}}, \ and\ \bibinfo
  {author} {\bibfnamefont {F.}~\bibnamefont {Schreiber}},\ }\href@noop {}
  {\bibfield  {journal} {\bibinfo  {journal} {Adv. Eng. Mater.}\ }\textbf
  {\bibinfo {volume} {11}},\ \bibinfo {pages} {291} (\bibinfo {year}
  {2009})}\BibitemShut {NoStop}%
\bibitem [{\citenamefont {Zhang}\ \emph {et~al.}(2009)\citenamefont {Zhang},
  \citenamefont {Barrena}, \citenamefont {Goswami}, \citenamefont {de~Oteyza},
  \citenamefont {Weis},\ and\ \citenamefont {Dosch}}]{Zhang_2009_PhysRevLett}%
  \BibitemOpen
  \bibfield  {author} {\bibinfo {author} {\bibfnamefont {X.}~\bibnamefont
  {Zhang}}, \bibinfo {author} {\bibfnamefont {E.}~\bibnamefont {Barrena}},
  \bibinfo {author} {\bibfnamefont {D.}~\bibnamefont {Goswami}}, \bibinfo
  {author} {\bibfnamefont {D.~G.}\ \bibnamefont {de~Oteyza}}, \bibinfo {author}
  {\bibfnamefont {C.}~\bibnamefont {Weis}}, \ and\ \bibinfo {author}
  {\bibfnamefont {H.}~\bibnamefont {Dosch}},\ }\href@noop {} {\bibfield
  {journal} {\bibinfo  {journal} {Phys. Rev. Lett.}\ }\textbf {\bibinfo
  {volume} {103}},\ \bibinfo {pages} {136101} (\bibinfo {year}
  {2009})}\BibitemShut {NoStop}%
\bibitem [{\citenamefont {D\"urr}\ \emph {et~al.}(2003)\citenamefont {D\"urr},
  \citenamefont {Schreiber}, \citenamefont {Ritley}, \citenamefont {Kruppa},
  \citenamefont {Krug}, \citenamefont {Dosch},\ and\ \citenamefont
  {Struth}}]{Durr_2003_PhysRevLett}%
  \BibitemOpen
  \bibfield  {author} {\bibinfo {author} {\bibfnamefont {A.~C.}\ \bibnamefont
  {D\"urr}}, \bibinfo {author} {\bibfnamefont {F.}~\bibnamefont {Schreiber}},
  \bibinfo {author} {\bibfnamefont {K.~A.}\ \bibnamefont {Ritley}}, \bibinfo
  {author} {\bibfnamefont {V.}~\bibnamefont {Kruppa}}, \bibinfo {author}
  {\bibfnamefont {J.}~\bibnamefont {Krug}}, \bibinfo {author} {\bibfnamefont
  {H.}~\bibnamefont {Dosch}}, \ and\ \bibinfo {author} {\bibfnamefont
  {B.}~\bibnamefont {Struth}},\ }\href@noop {} {\bibfield  {journal} {\bibinfo
  {journal} {Phys. Rev. Lett.}\ }\textbf {\bibinfo {volume} {90}},\ \bibinfo
  {pages} {016104} (\bibinfo {year} {2003})}\BibitemShut {NoStop}%
\bibitem [{\citenamefont {D\"urr}\ \emph {et~al.}(2006)\citenamefont {D\"urr},
  \citenamefont {Nickel}, \citenamefont {Sharma}, \citenamefont {T\"affner},\
  and\ \citenamefont {Dosch}}]{Durr_2005_ThinSolidFilms}%
  \BibitemOpen
  \bibfield  {author} {\bibinfo {author} {\bibfnamefont {A.}~\bibnamefont
  {D\"urr}}, \bibinfo {author} {\bibfnamefont {B.}~\bibnamefont {Nickel}},
  \bibinfo {author} {\bibfnamefont {V.}~\bibnamefont {Sharma}}, \bibinfo
  {author} {\bibfnamefont {U.}~\bibnamefont {T\"affner}}, \ and\ \bibinfo
  {author} {\bibfnamefont {H.}~\bibnamefont {Dosch}},\ }\href@noop {}
  {\bibfield  {journal} {\bibinfo  {journal} {Thin Solid Films}\ }\textbf
  {\bibinfo {volume} {503}},\ \bibinfo {pages} {127} (\bibinfo {year}
  {2006})}\BibitemShut {NoStop}%
\bibitem [{\citenamefont {Heinrich}\ \emph {et~al.}(2007)\citenamefont
  {Heinrich}, \citenamefont {Pflaum}, \citenamefont {Tripathi}, \citenamefont
  {Frey}, \citenamefont {Steigerwald},\ and\ \citenamefont
  {Siegrist}}]{HeinrichJPCC07}%
  \BibitemOpen
  \bibfield  {author} {\bibinfo {author} {\bibfnamefont {M.~A.}\ \bibnamefont
  {Heinrich}}, \bibinfo {author} {\bibfnamefont {J.}~\bibnamefont {Pflaum}},
  \bibinfo {author} {\bibfnamefont {A.~K.}\ \bibnamefont {Tripathi}}, \bibinfo
  {author} {\bibfnamefont {W.}~\bibnamefont {Frey}}, \bibinfo {author}
  {\bibfnamefont {M.~L.}\ \bibnamefont {Steigerwald}}, \ and\ \bibinfo {author}
  {\bibfnamefont {T.}~\bibnamefont {Siegrist}},\ }\href@noop {} {\bibfield
  {journal} {\bibinfo  {journal} {J. Phys. Chem. C}\ }\textbf {\bibinfo
  {volume} {111}},\ \bibinfo {pages} {18878} (\bibinfo {year}
  {2007})}\BibitemShut {NoStop}%
\bibitem [{\citenamefont {Kowarik}\ \emph {et~al.}(2006)\citenamefont
  {Kowarik}, \citenamefont {Gerlach}, \citenamefont {Sellner}, \citenamefont
  {Schreiber}, \citenamefont {Cavalcanti},\ and\ \citenamefont
  {Konovalov}}]{Kowarik_2006_PhysRevLett}%
  \BibitemOpen
  \bibfield  {author} {\bibinfo {author} {\bibfnamefont {S.}~\bibnamefont
  {Kowarik}}, \bibinfo {author} {\bibfnamefont {A.}~\bibnamefont {Gerlach}},
  \bibinfo {author} {\bibfnamefont {S.}~\bibnamefont {Sellner}}, \bibinfo
  {author} {\bibfnamefont {F.}~\bibnamefont {Schreiber}}, \bibinfo {author}
  {\bibfnamefont {L.}~\bibnamefont {Cavalcanti}}, \ and\ \bibinfo {author}
  {\bibfnamefont {O.}~\bibnamefont {Konovalov}},\ }\href@noop {} {\bibfield
  {journal} {\bibinfo  {journal} {Phys. Rev. Lett.}\ }\textbf {\bibinfo
  {volume} {96}},\ \bibinfo {eid} {125504} (\bibinfo {year}
  {2006})}\BibitemShut {NoStop}%
\bibitem [{\citenamefont {Banerjee}\ \emph {et~al.}(2013)\citenamefont
  {Banerjee}, \citenamefont {Nov{\'a}k}, \citenamefont {Frank}, \citenamefont
  {Lorch}, \citenamefont {Hinderhofer}, \citenamefont {Gerlach},\ and\
  \citenamefont {Schreiber}}]{Banerjee_2013_PhysRevLett}%
  \BibitemOpen
  \bibfield  {author} {\bibinfo {author} {\bibfnamefont {R.}~\bibnamefont
  {Banerjee}}, \bibinfo {author} {\bibfnamefont {J.}~\bibnamefont {Nov{\'a}k}},
  \bibinfo {author} {\bibfnamefont {C.}~\bibnamefont {Frank}}, \bibinfo
  {author} {\bibfnamefont {C.}~\bibnamefont {Lorch}}, \bibinfo {author}
  {\bibfnamefont {A.}~\bibnamefont {Hinderhofer}}, \bibinfo {author}
  {\bibfnamefont {A.}~\bibnamefont {Gerlach}}, \ and\ \bibinfo {author}
  {\bibfnamefont {F.}~\bibnamefont {Schreiber}},\ }\href@noop {} {\bibfield
  {journal} {\bibinfo  {journal} {Phys. Rev. Lett.}\ }\textbf {\bibinfo
  {volume} {110}},\ \bibinfo {pages} {185506} (\bibinfo {year}
  {2013})}\BibitemShut {NoStop}%
\bibitem [{\citenamefont {Aufderheide}\ \emph {et~al.}(2012)\citenamefont
  {Aufderheide}, \citenamefont {Broch}, \citenamefont {Nov\'ak}, \citenamefont
  {Hinderhofer}, \citenamefont {Nervo}, \citenamefont {Gerlach}, \citenamefont
  {Banerjee},\ and\ \citenamefont {Schreiber}}]{Aufderheide_2012_PhysRevLett}%
  \BibitemOpen
  \bibfield  {author} {\bibinfo {author} {\bibfnamefont {A.}~\bibnamefont
  {Aufderheide}}, \bibinfo {author} {\bibfnamefont {K.}~\bibnamefont {Broch}},
  \bibinfo {author} {\bibfnamefont {J.}~\bibnamefont {Nov\'ak}}, \bibinfo
  {author} {\bibfnamefont {A.}~\bibnamefont {Hinderhofer}}, \bibinfo {author}
  {\bibfnamefont {R.}~\bibnamefont {Nervo}}, \bibinfo {author} {\bibfnamefont
  {A.}~\bibnamefont {Gerlach}}, \bibinfo {author} {\bibfnamefont
  {R.}~\bibnamefont {Banerjee}}, \ and\ \bibinfo {author} {\bibfnamefont
  {F.}~\bibnamefont {Schreiber}},\ }\href@noop {} {\bibfield  {journal}
  {\bibinfo  {journal} {Phys. Rev. Lett.}\ }\textbf {\bibinfo {volume} {109}},\
  \bibinfo {pages} {156102} (\bibinfo {year} {2012})}\BibitemShut {NoStop}%
\bibitem [{\citenamefont {Revenant}\ \emph {et~al.}(2004)\citenamefont
  {Revenant}, \citenamefont {Leroy}, \citenamefont {Lazzari}, \citenamefont
  {Renaud},\ and\ \citenamefont {Henry}}]{Revenant_PRB_2004}%
  \BibitemOpen
  \bibfield  {author} {\bibinfo {author} {\bibfnamefont {C.}~\bibnamefont
  {Revenant}}, \bibinfo {author} {\bibfnamefont {F.}~\bibnamefont {Leroy}},
  \bibinfo {author} {\bibfnamefont {R.}~\bibnamefont {Lazzari}}, \bibinfo
  {author} {\bibfnamefont {G.}~\bibnamefont {Renaud}}, \ and\ \bibinfo {author}
  {\bibfnamefont {C. R.}~\bibnamefont {Henry}},\ }\href@noop {} {\bibfield
  {journal} {\bibinfo  {journal} {Phys. Rev. B}\ }\textbf {\bibinfo {volume}
  {69}},\ \bibinfo {pages} {035411} (\bibinfo {year} {2004})}\BibitemShut
  {NoStop}%
\bibitem [{\citenamefont {Daillant}\ and\ \citenamefont
  {Gibaud}(2009)}]{Daillant_2009_book}%
  \BibitemOpen
  \bibfield  {author} {\bibinfo {author} {\bibfnamefont {J.}~\bibnamefont
  {Daillant}}\ and\ \bibinfo {author} {\bibfnamefont {A.}~\bibnamefont
  {Gibaud}},\ }\href@noop {} {\emph {\bibinfo {title} {X-Ray and Neutron
  Reflectivity}}}\ (\bibinfo  {publisher} {Springer},\ \bibinfo {address}
  {Paris},\ \bibinfo {year} {2009})\BibitemShut {NoStop}%
\bibitem [{\citenamefont {Ritley}\ \emph {et~al.}(2001)\citenamefont {Ritley},
  \citenamefont {Krause}, \citenamefont {Schreiber},\ and\ \citenamefont
  {Dosch}}]{Ritley_2001_RevSciInstrum}%
  \BibitemOpen
  \bibfield  {author} {\bibinfo {author} {\bibfnamefont {K.~A.}\ \bibnamefont
  {Ritley}}, \bibinfo {author} {\bibfnamefont {B.}~\bibnamefont {Krause}},
  \bibinfo {author} {\bibfnamefont {F.}~\bibnamefont {Schreiber}}, \ and\
  \bibinfo {author} {\bibfnamefont {H.}~\bibnamefont {Dosch}},\ }\href@noop {}
  {\bibfield  {journal} {\bibinfo  {journal} {Rev. Sci. Instrum.}\ }\textbf
  {\bibinfo {volume} {72}},\ \bibinfo {pages} {1453} (\bibinfo {year}
  {2001})}\BibitemShut {NoStop}%
\bibitem [{\citenamefont {Witte}\ and\ \citenamefont
  {W\"oll}(2004)}]{Witte_2004_JMaterRes}%
  \BibitemOpen
  \bibfield  {author} {\bibinfo {author} {\bibfnamefont {G.}~\bibnamefont
  {Witte}}\ and\ \bibinfo {author} {\bibfnamefont {C.}~\bibnamefont {W\"oll}},\
  }\href {\doibase 10.1557/JMR.2004.0251} {\bibfield  {journal} {\bibinfo
  {journal} {J. Mater. Res.}\ }\textbf {\bibinfo {volume} {19}},\ \bibinfo
  {pages} {1889} (\bibinfo {year} {2004})}\BibitemShut {NoStop}%
\bibitem [{\citenamefont {Schreiber}(2004)}]{Schreiber_2004_Physstatsola}%
  \BibitemOpen
  \bibfield  {author} {\bibinfo {author} {\bibfnamefont {F.}~\bibnamefont
  {Schreiber}},\ }\href {\doibase 10.1002/pssa.200404334} {\bibfield  {journal}
  {\bibinfo  {journal} {Phys. stat. sol. (a)}\ }\textbf {\bibinfo {volume}
  {201}},\ \bibinfo {pages} {1037} (\bibinfo {year} {2004})}\BibitemShut
  {NoStop}%
\bibitem [{\citenamefont {Hinderhofer}\ and\ \citenamefont
  {Schreiber}(2012)}]{Hinderhofer_2012_ChemPhysChem}%
  \BibitemOpen
  \bibfield  {author} {\bibinfo {author} {\bibfnamefont {A.}~\bibnamefont
  {Hinderhofer}}\ and\ \bibinfo {author} {\bibfnamefont {F.}~\bibnamefont
  {Schreiber}},\ }\href@noop {} {\bibfield  {journal} {\bibinfo  {journal}
  {ChemPhysChem}\ }\textbf {\bibinfo {volume} {13}},\ \bibinfo {pages} {628}
  (\bibinfo {year} {2012})}\BibitemShut {NoStop}%
\bibitem [{\citenamefont {Roth}\ \emph {et~al.}(2011)\citenamefont {Roth},
  \citenamefont {Herzog}, \citenamefont {K\"{o}rstgens}, \citenamefont
  {Buffet}, \citenamefont {Schwartzkopf}, \citenamefont {Perlich},
  \citenamefont {Abul~Kashem}, \citenamefont {D\"{o}hrmann}, \citenamefont
  {Gehrke}, \citenamefont {Rothkirch}, \citenamefont {Stassig}, \citenamefont
  {Wurth}, \citenamefont {Benecke}, \citenamefont {Li}, \citenamefont {Fratzl},
  \citenamefont {Rawolle},\ and\ \citenamefont
  {M\"{u}ller-Buschbaum}}]{Roth_JPhysCondensMatter_2011}%
  \BibitemOpen
  \bibfield  {author} {\bibinfo {author} {\bibfnamefont {S.~V.}\ \bibnamefont
  {Roth}}, \bibinfo {author} {\bibfnamefont {G.}~\bibnamefont {Herzog}},
  \bibinfo {author} {\bibfnamefont {V.}~\bibnamefont {K\"{o}rstgens}}, \bibinfo
  {author} {\bibfnamefont {A.}~\bibnamefont {Buffet}}, \bibinfo {author}
  {\bibfnamefont {M.}~\bibnamefont {Schwartzkopf}}, \bibinfo {author}
  {\bibfnamefont {J.}~\bibnamefont {Perlich}}, \bibinfo {author} {\bibfnamefont
  {M.~M.}\ \bibnamefont {Abul~Kashem}}, \bibinfo {author} {\bibfnamefont
  {R.}~\bibnamefont {D\"{o}hrmann}}, \bibinfo {author} {\bibfnamefont
  {R.}~\bibnamefont {Gehrke}}, \bibinfo {author} {\bibfnamefont
  {A.}~\bibnamefont {Rothkirch}}, \bibinfo {author} {\bibfnamefont
  {K.}~\bibnamefont {Stassig}}, \bibinfo {author} {\bibfnamefont
  {W.}~\bibnamefont {Wurth}}, \bibinfo {author} {\bibfnamefont
  {G.}~\bibnamefont {Benecke}}, \bibinfo {author} {\bibfnamefont
  {C.}~\bibnamefont {Li}}, \bibinfo {author} {\bibfnamefont {P.}~\bibnamefont
  {Fratzl}}, \bibinfo {author} {\bibfnamefont {M.}~\bibnamefont {Rawolle}}, \
  and\ \bibinfo {author} {\bibfnamefont {P.}~\bibnamefont
  {M\"{u}ller-Buschbaum}},\ }\href@noop {} {\bibfield  {journal} {\bibinfo
  {journal} {J. Phys. Condens. Matter}\ }\textbf {\bibinfo {volume} {23}},\
  \bibinfo {pages} {254208} (\bibinfo {year} {2011})}\BibitemShut {NoStop}%
\bibitem [{\citenamefont {Buffet}\ \emph {et~al.}(2012)\citenamefont {Buffet},
  \citenamefont {Rothkirch}, \citenamefont {D\"{o}hrmann}, \citenamefont
  {K\"{o}rstgens}, \citenamefont {Abul~Kashem}, \citenamefont {Perlich},
  \citenamefont {Herzog}, \citenamefont {Schwartzkopf}, \citenamefont {Gehrke},
  \citenamefont {M\"{u}ller-Buschbaum},\ and\ \citenamefont
  {Roth}}]{Buffet_JSynRad_2012}%
  \BibitemOpen
  \bibfield  {author} {\bibinfo {author} {\bibfnamefont {A.}~\bibnamefont
  {Buffet}}, \bibinfo {author} {\bibfnamefont {A.}~\bibnamefont {Rothkirch}},
  \bibinfo {author} {\bibfnamefont {R.}~\bibnamefont {D\"{o}hrmann}}, \bibinfo
  {author} {\bibfnamefont {V.}~\bibnamefont {K\"{o}rstgens}}, \bibinfo {author}
  {\bibfnamefont {M.~M.}\ \bibnamefont {Abul~Kashem}}, \bibinfo {author}
  {\bibfnamefont {J.}~\bibnamefont {Perlich}}, \bibinfo {author} {\bibfnamefont
  {G.}~\bibnamefont {Herzog}}, \bibinfo {author} {\bibfnamefont
  {M.}~\bibnamefont {Schwartzkopf}}, \bibinfo {author} {\bibfnamefont
  {R.}~\bibnamefont {Gehrke}}, \bibinfo {author} {\bibfnamefont
  {P.}~\bibnamefont {M\"{u}ller-Buschbaum}}, \ and\ \bibinfo {author}
  {\bibfnamefont {S.~V.}\ \bibnamefont {Roth}},\ }\href@noop {} {\bibfield
  {journal} {\bibinfo  {journal} {J. Synchrotron Rad.}\ }\textbf {\bibinfo
  {volume} {19}},\ \bibinfo {pages} {647} (\bibinfo {year} {2012})}\BibitemShut
  {NoStop}%
\bibitem [{\citenamefont {Santoro}\ \emph
  {et~al.}(2014{\natexlab{b}})\citenamefont {Santoro}, \citenamefont {Buffet},
  \citenamefont {D\"{o}hrmann}, \citenamefont {Yu}, \citenamefont
  {K\"{o}rstgens}, \citenamefont {M\"{u}ller-Buschbaum}, \citenamefont {Gedde},
  \citenamefont {Hedenqvist},\ and\ \citenamefont
  {Roth}}]{Santoro_RevSciInstrum_2014}%
  \BibitemOpen
  \bibfield  {author} {\bibinfo {author} {\bibfnamefont {G.}~\bibnamefont
  {Santoro}}, \bibinfo {author} {\bibfnamefont {A.}~\bibnamefont {Buffet}},
  \bibinfo {author} {\bibfnamefont {R.}~\bibnamefont {D\"{o}hrmann}}, \bibinfo
  {author} {\bibfnamefont {S.}~\bibnamefont {Yu}}, \bibinfo {author}
  {\bibfnamefont {V.}~\bibnamefont {K\"{o}rstgens}}, \bibinfo {author}
  {\bibfnamefont {P.}~\bibnamefont {M\"{u}ller-Buschbaum}}, \bibinfo {author}
  {\bibfnamefont {U.}~\bibnamefont {Gedde}}, \bibinfo {author} {\bibfnamefont
  {M.}~\bibnamefont {Hedenqvist}}, \ and\ \bibinfo {author} {\bibfnamefont
  {S.~V.}\ \bibnamefont {Roth}},\ }\href@noop {} {\bibfield  {journal}
  {\bibinfo  {journal} {Rev. Sci. Instrum.}\ }\textbf {\bibinfo {volume}
  {85}},\ \bibinfo {pages} {043901} (\bibinfo {year}
  {2014}{\natexlab{b}})}\BibitemShut {NoStop}%
\bibitem [{\citenamefont {Ne\v{c}as}\ and\ \citenamefont
  {Klapetek}(2012)}]{Necas_2011_centreurjphys}%
  \BibitemOpen
  \bibfield  {author} {\bibinfo {author} {\bibfnamefont {D.}~\bibnamefont
  {Ne\v{c}as}}\ and\ \bibinfo {author} {\bibfnamefont {P.}~\bibnamefont
  {Klapetek}},\ }\href@noop {} {\bibfield  {journal} {\bibinfo  {journal}
  {Centr. Eur. J. Phys.}\ }\textbf {\bibinfo {volume} {10}},\ \bibinfo {pages}
  {181} (\bibinfo {year} {2012})}\BibitemShut {NoStop}%
\bibitem [{\citenamefont {D\"{u}rr}\ \emph {et~al.}(2003)\citenamefont
  {D\"{u}rr}, \citenamefont {Koch}, \citenamefont {Kelsch}, \citenamefont
  {R\"{u}hm}, \citenamefont {Ghijsen}, \citenamefont {Johnson}, \citenamefont
  {Pireaux}, \citenamefont {Schwartz}, \citenamefont {Schreiber}, \citenamefont
  {Dosch},\ and\ \citenamefont {Kahn}}]{Durr_2003_PRB}%
  \BibitemOpen
  \bibfield  {author} {\bibinfo {author} {\bibfnamefont {A.~C.}\ \bibnamefont
  {D\"{u}rr}}, \bibinfo {author} {\bibfnamefont {N.}~\bibnamefont {Koch}},
  \bibinfo {author} {\bibfnamefont {M.}~\bibnamefont {Kelsch}}, \bibinfo
  {author} {\bibfnamefont {A.}~\bibnamefont {R\"{u}hm}}, \bibinfo {author}
  {\bibfnamefont {J.}~\bibnamefont {Ghijsen}}, \bibinfo {author} {\bibfnamefont
  {R.~L.}\ \bibnamefont {Johnson}}, \bibinfo {author} {\bibfnamefont {J.-J.}\
  \bibnamefont {Pireaux}}, \bibinfo {author} {\bibfnamefont {J.}~\bibnamefont
  {Schwartz}}, \bibinfo {author} {\bibfnamefont {F.}~\bibnamefont {Schreiber}},
  \bibinfo {author} {\bibfnamefont {H.}~\bibnamefont {Dosch}}, \ and\ \bibinfo
  {author} {\bibfnamefont {A.}~\bibnamefont {Kahn}},\ }\href@noop {} {\bibfield
   {journal} {\bibinfo  {journal} {Phys. Rev. B}\ }\textbf {\bibinfo {volume}
  {68}},\ \bibinfo {pages} {115428} (\bibinfo {year} {2003})}\BibitemShut
  {NoStop}%
\bibitem [{\citenamefont {D\"{u}rr}\ \emph {et~al.}(2002)\citenamefont
  {D\"{u}rr}, \citenamefont {Schreiber}, \citenamefont {M\"{u}nch},
  \citenamefont {Karl}, \citenamefont {Krause}, \citenamefont {Kruppa},\ and\
  \citenamefont {Dosch}}]{Durr_2002_APL}%
  \BibitemOpen
  \bibfield  {author} {\bibinfo {author} {\bibfnamefont {A.~C.}\ \bibnamefont
  {D\"{u}rr}}, \bibinfo {author} {\bibfnamefont {F.}~\bibnamefont {Schreiber}},
  \bibinfo {author} {\bibfnamefont {M.}~\bibnamefont {M\"{u}nch}}, \bibinfo
  {author} {\bibfnamefont {N.}~\bibnamefont {Karl}}, \bibinfo {author}
  {\bibfnamefont {B.}~\bibnamefont {Krause}}, \bibinfo {author} {\bibfnamefont
  {V.}~\bibnamefont {Kruppa}}, \ and\ \bibinfo {author} {\bibfnamefont
  {H.}~\bibnamefont {Dosch}},\ }\href@noop {} {\bibfield  {journal} {\bibinfo
  {journal} {Appl. Phys. Lett.}\ }\textbf {\bibinfo {volume} {81}},\ \bibinfo
  {pages} {2276} (\bibinfo {year} {2002})}\BibitemShut {NoStop}%
\bibitem [{\citenamefont {Babonneau}(2010)}]{Babonneau_2010_JApplCrystallogr}%
  \BibitemOpen
  \bibfield  {author} {\bibinfo {author} {\bibfnamefont {D.}~\bibnamefont
  {Babonneau}},\ }\href@noop {} {\bibfield  {journal} {\bibinfo  {journal} {J.
  Appl. Crystallogr.}\ }\textbf {\bibinfo {volume} {43}},\ \bibinfo {pages}
  {929} (\bibinfo {year} {2010})}\BibitemShut {NoStop}%
\bibitem [{\citenamefont {Lee}\ \emph {et~al.}(2003)\citenamefont {Lee},
  \citenamefont {Chu}, \citenamefont {Choi}, \citenamefont {Lang},
  \citenamefont {Srajer}, \citenamefont {Sinha}, \citenamefont {Metlushko},\
  and\ \citenamefont {Ilic}}]{Lee_2003_ApplPhysLett}%
  \BibitemOpen
  \bibfield  {author} {\bibinfo {author} {\bibfnamefont {D.~R.}\ \bibnamefont
  {Lee}}, \bibinfo {author} {\bibfnamefont {Y.~S.}\ \bibnamefont {Chu}},
  \bibinfo {author} {\bibfnamefont {Y.}~\bibnamefont {Choi}}, \bibinfo {author}
  {\bibfnamefont {J.~C.}\ \bibnamefont {Lang}}, \bibinfo {author}
  {\bibfnamefont {G.}~\bibnamefont {Srajer}}, \bibinfo {author} {\bibfnamefont
  {S.~K.}\ \bibnamefont {Sinha}}, \bibinfo {author} {\bibfnamefont
  {V.}~\bibnamefont {Metlushko}}, \ and\ \bibinfo {author} {\bibfnamefont
  {B.}~\bibnamefont {Ilic}},\ }\href@noop {} {\bibfield  {journal} {\bibinfo
  {journal} {Appl. Phys. Lett.}\ }\textbf {\bibinfo {volume} {82}},\ \bibinfo
  {pages} {982} (\bibinfo {year} {2003})}\BibitemShut {NoStop}%
\bibitem [{\citenamefont {Basu}\ and\ \citenamefont
  {Sanyal}(2002)}]{Basu_phys_rep_2002}%
  \BibitemOpen
  \bibfield  {author} {\bibinfo {author} {\bibfnamefont {J.~K.}\ \bibnamefont
  {Basu}}\ and\ \bibinfo {author} {\bibfnamefont {M.~K.}\ \bibnamefont
  {Sanyal}},\ }\href@noop {} {\bibfield  {journal} {\bibinfo  {journal}
  {Physics Reports}\ }\textbf {\bibinfo {volume} {363}},\ \bibinfo {pages} {1 }
  (\bibinfo {year} {2002})}\BibitemShut {NoStop}%
\bibitem [{\citenamefont {Pietsch}, \citenamefont {Hol\'{y}},\ and\
  \citenamefont {Baumbach}(2004)}]{Pietsch_2004_book}%
  \BibitemOpen
  \bibfield  {author} {\bibinfo {author} {\bibfnamefont {U.}~\bibnamefont
  {Pietsch}}, \bibinfo {author} {\bibfnamefont {V.}~\bibnamefont {Hol\'{y}}}, \
  and\ \bibinfo {author} {\bibfnamefont {T.}~\bibnamefont {Baumbach}},\
  }\href@noop {} {\emph {\bibinfo {title} {High-Resolution X-Ray Scattering:
  From Thin Films to Lateral Nanostructures}}}\ (\bibinfo  {publisher}
  {Springer-Verlag},\ \bibinfo {year} {2004})\BibitemShut {NoStop}%
\bibitem [{\citenamefont {Hansen}\ and\ \citenamefont
  {McDonald}(2006)}]{Hansen_2006_book}%
  \BibitemOpen
  \bibfield  {author} {\bibinfo {author} {\bibfnamefont {J.~P.}\ \bibnamefont
  {Hansen}}\ and\ \bibinfo {author} {\bibfnamefont {I.~R.}\ \bibnamefont
  {McDonald}},\ }\href@noop {} {\emph {\bibinfo {title} {Theory of simple
  liquids}}},\ \bibinfo {edition} {3rd}\ ed.\ (\bibinfo  {publisher} {Academic
  Press},\ \bibinfo {address} {New York},\ \bibinfo {year} {2006})\BibitemShut
  {NoStop}%
\bibitem [{\citenamefont {Guinier}(1994)}]{guinier_1994}%
  \BibitemOpen
  \bibfield  {author} {\bibinfo {author} {\bibfnamefont {A.}~\bibnamefont
  {Guinier}},\ }\href@noop {} {\emph {\bibinfo {title} {{X-Ray Diffraction in
  Crystals, Imperfect Crystals, and Amorphous Bodies}}}}\ (\bibinfo
  {publisher} {Dover Publications, Inc.},\ \bibinfo {address} {New York},\
  \bibinfo {year} {1994})\BibitemShut {NoStop}%
\bibitem [{\citenamefont {Lazzari}(2002)}]{Lazzari}%
  \BibitemOpen
  \bibfield  {author} {\bibinfo {author} {\bibfnamefont {R.}~\bibnamefont
  {Lazzari}},\ }\href@noop {} {\bibfield  {journal} {\bibinfo  {journal} {J.
  Appl. Cryst.}\ }\textbf {\bibinfo {volume} {35}},\ \bibinfo {pages} {406}
  (\bibinfo {year} {2002})}\BibitemShut {NoStop}%
\bibitem [{\citenamefont {Renaud}, \citenamefont {Lazzari},\ and\ \citenamefont
  {Leroy}(2009)}]{Renaud_2009_SurfSciRep}%
  \BibitemOpen
  \bibfield  {author} {\bibinfo {author} {\bibfnamefont {G.}~\bibnamefont
  {Renaud}}, \bibinfo {author} {\bibfnamefont {R.}~\bibnamefont {Lazzari}}, \
  and\ \bibinfo {author} {\bibfnamefont {F.}~\bibnamefont {Leroy}},\
  }\href@noop {} {\bibfield  {journal} {\bibinfo  {journal} {Surf. Sci. Rep.}\
  }\textbf {\bibinfo {volume} {64}},\ \bibinfo {pages} {255} (\bibinfo {year}
  {2009})}\BibitemShut {NoStop}%
\bibitem [{\citenamefont {Shannon}(1948)}]{Shannon_1948}%
  \BibitemOpen
  \bibfield  {author} {\bibinfo {author} {\bibfnamefont {C.~E.}\ \bibnamefont
  {Shannon}},\ }\href@noop {} {\bibfield  {journal} {\bibinfo  {journal} {Bell
  Syst. Tech. J.}\ }\textbf {\bibinfo {volume} {27}},\ \bibinfo {pages} {379}
  (\bibinfo {year} {1948})}\BibitemShut {NoStop}%
\bibitem [{\citenamefont {Hamilton}(2000)}]{Hamilton_2000}%
  \BibitemOpen
  \bibfield  {author} {\bibinfo {author} {\bibfnamefont {A.~J.~S.}\
  \bibnamefont {Hamilton}},\ }\href@noop {} {\bibfield  {journal} {\bibinfo
  {journal} {Mon. Not. R. Astron. Soc.}\ }\textbf {\bibinfo {volume} {312}},\
  \bibinfo {pages} {257} (\bibinfo {year} {2000})}\BibitemShut {NoStop}%
\bibitem [{\citenamefont {Botan}\ \emph {et~al.}(2009)\citenamefont {Botan},
  \citenamefont {Pesth}, \citenamefont {Schilling},\ and\ \citenamefont
  {Oettel}}]{Botan_2000_PRE}%
  \BibitemOpen
  \bibfield  {author} {\bibinfo {author} {\bibfnamefont {V.}~\bibnamefont
  {Botan}}, \bibinfo {author} {\bibfnamefont {F.}~\bibnamefont {Pesth}},
  \bibinfo {author} {\bibfnamefont {T.}~\bibnamefont {Schilling}}, \ and\
  \bibinfo {author} {\bibfnamefont {M.}~\bibnamefont {Oettel}},\ }\href@noop {}
  {\bibfield  {journal} {\bibinfo  {journal} {Phys. Rev. E}\ }\textbf {\bibinfo
  {volume} {79}},\ \bibinfo {pages} {061402} (\bibinfo {year}
  {2009})}\BibitemShut {NoStop}%
\bibitem [{\citenamefont {MATLAB}(2012)}]{MATLAB:2012}%
  \BibitemOpen
  \bibfield  {author} {\bibinfo {author} {\bibnamefont {MATLAB}},\ }\href@noop
  {} {\emph {\bibinfo {title} {version 8.0.0.783 (R2012b)}}}\ (\bibinfo
  {publisher} {The MathWorks Inc.},\ \bibinfo {address} {Natick,
  Massachusetts},\ \bibinfo {year} {2012})\BibitemShut {NoStop}%
\bibitem [{\citenamefont {Mfit}(2005)}]{Mfit:2005}%
  \BibitemOpen
  \bibfield  {author} {\bibinfo {author} {\bibnamefont {Mfit}},\ }\href@noop {}
  {\emph {\bibinfo {title} {version 4.3.3}}}\ (\bibinfo  {publisher} {Insitut
  Laue-Langevin},\ \bibinfo {address} {Grenoble},\ \bibinfo {year}
  {2005})\BibitemShut {NoStop}%
\bibitem [{\citenamefont {Nelder}\ and\ \citenamefont
  {Mead}(1965)}]{Nelder_1965}%
  \BibitemOpen
  \bibfield  {author} {\bibinfo {author} {\bibfnamefont {J.~A.}\ \bibnamefont
  {Nelder}}\ and\ \bibinfo {author} {\bibfnamefont {R.}~\bibnamefont {Mead}},\
  }\href@noop {} {\bibfield  {journal} {\bibinfo  {journal} {Comput. J.}\
  }\textbf {\bibinfo {volume} {7}},\ \bibinfo {pages} {308} (\bibinfo {year}
  {1965})}\BibitemShut {NoStop}%
\end{thebibliography}
%

\end{document}